\documentclass[aip,jap,numerical,amssymb,
preprint,
reprint,
unsortedaddress
]{revtex4-1}

\usepackage{graphicx}
\usepackage{amsmath}

\begin{document}

\title{Cylindrical coordinate representation for multiband Hamiltonians}
\author{Eduard Takhtamirov}
\affiliation{Conestoga College Institute of Technology and Advanced Learning, Kitchener, Ontario N2G 4M4, Canada}
\date{\today}

\begin{abstract}
Rotationally invariant combinations of the Brillouin zone-center Bloch functions are used as basis function to express in cylindrical coordinates the valence-band and Kane envelope-function Hamiltonians for wurtzite and zinc-blende semiconductor heterostructures. For cylindrically symmetric systems, this basis allows to treat the envelope functions as eigenstates of the operator of projection of total angular momentum on the symmetry axis, with the operator's eigenvalue conventionally entering the Hamiltonians as a parameter. Complementing the Hamiltonians with boundary conditions for the envelope functions on the symmetry axis, we present for the first time a complete formalism for efficient modeling and description of multiband electron states in low-dimensional semiconductor structures with cylindrical symmetry. To demonstrate the potency of the cylindrical symmetry approximation and establish a criterion of its applicability for actual structures, we map the ground and several excited valence-band states in an isolated wurtzite GaN quantum wire of a hexagonal cross-section to the states in an equivalent quantum wire of a circular cross-section.
\end{abstract}

\maketitle

\section{Introduction}\label{intro}

Artificial electron systems with 2D and 3D quantum confinement, realized in semiconductor quantum wires (QWs) and quantum dots (QDs), have become a subject of extensive study in the last two decades. \cite{QD} Advances in nanotechnology guarantee that QWs and QDs will eventually become a basis for future-generation photonic and electronic devices, \cite{Kapon,Petroff,Gudiksen,Anikeeva} with QDs being viewed as one of the most promising candidates for use in solid-state quantum computation. \cite{Hawrylak_QC,Loss,Elzerman,QC_holes} Thus, the problem of description of electron states in such systems is actual.

An impressive progress in the application of atomistic approaches to real nanostructures \cite{Zunger_qd,DiCarlo,Persson,Mourad} has not shattered the positions of the envelope function (EF) method, which remains the most convenient means to study fundamental properties of electron states and virtually the only efficient and flexible tool for optimization of nanostructure-based devices. A number of shortcomings of the original method, \cite{Harrison,Zunger-far} associated with the behavior of the wave function at heterointerfaces, have been proved not to be critical when recipes of the method's modification have been developed and applied. \cite{Ivchenko_GX,Krebs,Ivchenko,Takhtamirov,gamma-x,stripes} Furthermore, for typical relatively large quantum mechanical systems, for which the ratio of the number of heterointerface atoms to the number of atoms located inside homogeneous materials is small, the interface effects are usually weak, and the applicability of even the classic effective-mass (EM) approximation \cite{Luttinger,Luttinger2,Leibler} can be justified. \cite{Zunger_large_good,Marquardt}

Already in the framework of the EM approximation, the calculation of electron states in QWs and especially in QDs is still an arduous and resource-demanding task due to the necessity to solve a set of Schr\"odinger-type partial differential equations, yet the problem is usually complicated by the strain and electromechanical subtasks. \cite{Grundmann,Stier,Andreev} Fortunately, for electron systems that can be approximated as cylindrically or spherically symmetric, the analysis can be significantly facilitated by the reduction in the number of independent variables. For a single-band EM equation, which is analogous to the usual Schr\"odinger equation, the corresponding transformation to cylindrical or spherical coordinates is a textbook task. \cite{Messiah,Landavshitz-QM} The same direct transformation does not, however, result in explicit radially symmetric EM equations to describe electron states in complex bands. A different approach is required in this case.

Such an approach has been presented by Sercel and Vahala \cite{SV} for zinc-blende materials in the spherical band approximation. \cite{Baldereschi} They used the standard basis of the Brillouin zone-center Bloch functions as eigenstates of the operator of total angular momentum, \cite{Luttinger,Luttinger2} employing a formal analogy of the total wave function, which is a product of the zone-center Bloch functions and EFs, with the wave function of a two-particle system. Here, we develop and apply an alternative representation on a basis of rotationally invariant combinations of the Brillouin zone-center Bloch functions. This basis allows to directly transfer the symmetry properties of the total wave function to the corresponding EFs. We consider cylindrically symmetric electron systems, for which the EFs can thus be chosen as eigenstates of the operator of projection of total angular momentum on the symmetry axis.

The paper is organized as follows. In Sec.~\ref{II_valence_band} we introduce the notations for the valence-band Hamiltonian for wurtzite heterostructures in Cartesian coordinates, make a cylindrical coordinate transformation, derive the boundary conditions (BCs) for the EFs on the symmetry axis, and specify how the results should be adapted for zinc-blende structures. In Sec.~\ref{III_Kane} we generalize the results of Sec.~\ref{II_valence_band} for the Kane model. \cite{Kane} As an application of the results, in Sec.~\ref{IV_holes} we study isolated [0001] wurtzite GaN QWs with the hexagonal and circular cross-sections. We compare the spectra of the valence-band states for these two cases. In Sec.~\ref{V_discuss} we argue that our method is not merely a good alternative to the cumbersome formalism of Sercel and Vahala, but also the only complete tool for efficient modeling of multiband electron states in cylindrically symmetric systems. We also discuss the radial symmetry approximation for actual heterostructures. Conclusions are given in Sec.~\ref{V_concl}.

\section{Hamiltonian for valence band}\label{II_valence_band}

\subsection{Hamiltonian for wurtzite in Cartesian coordinates}\label{II_wurtzite}

In this subsection we introduce the notations necessary for the following. We deal with the steady-state Schr\"odinger equation in the EM approximation for valence-band states in wurtzite heterostructures, \cite{Bir-Pikus,Chuang,Sirenko,Ren}
\begin{equation}\label{H}
\mathbf H \mathbf F \equiv \left( \mathbf H^{(0)} + \mathbf H^{(\sigma)} + \mathbf H^{(k)} + \mathbf H^{(\varepsilon)} \right) \mathbf F = E \mathbf F,
\end{equation}
with an eigenenergy $E$. The equation is constructed using the orthonormal Brillouin zone-center Bloch functions of the reference [potential well] semiconductor, which are written as $\mathbf u = \left( u_x\ u_y\ u_z\right)$ in the matrix form. They can be chosen real. In Cartesian coordinates $\mathbf r = (x,y)$ and $z$, the functions $u_x = u_x \left(\mathbf r ,z\right)$ and $u_y= u_y \left(\mathbf r , z \right)$ transform as the coordinates $x$ and $y$ belonging to the representation $\Gamma_6$ of the space group $C_{6v}$, and the function $u_z= u_z \left(\mathbf r ,z \right)$ transforms as the coordinate $z$, along the $c$-axis of wurtzite, belonging to the representation $\Gamma_1$.
The EF matrix $\mathbf F = \mathbf F \left( \mathbf r , z\right)$ has three components, $\mathbf F = \left(F_x\ F_y\ F_z \right)^{\rm T}$, where the symbol ${\rm T}$ stands for the transpose of a matrix or an operator, so that the total wave function $\Psi = \Psi \left( \mathbf r, z\right)$ is expressed as follows:
\begin{equation} \label{psi}
\Psi = \sum_{j=x,y,z} u_j F_j = \mathbf u \mathbf F.
\end{equation}
The basis functions $\mathbf u$ are spinless, \cite{Luttinger} but each EF component $F_j$ is a spinor with two elements,
\begin{equation}
F_j =
\begin{pmatrix}
F^{(u)}_j\\
F^{(d)}_j
\end{pmatrix}, \quad j=x,y,z.
\end{equation}

In Eq.~(\ref{H}), the Hamiltonian $\mathbf H^{(0)}$ represents the potential energy of an electron,
\begin{equation}
\mathbf H^{(0)} =
\begin{pmatrix}
U_{v6}&0&0\cr
0&U_{v6}&0\cr
0&0&U_{v1}
\end{pmatrix},\label{H0}
\end{equation}
where $U_{v6} = U_{v6}\left(\mathbf r, z\right)$ and $U_{v1} = U_{v1}\left(\mathbf r, z\right)$ are the position-dependent edges of the valence bands $\Gamma_6$ and $\Gamma_1$, respectively, which include an external scalar potential present.

The Hamiltonian $\mathbf H^{(\sigma)}$ defines the spin-orbit interaction, which is taken into account in the first-order perturbation theory, neglecting small terms linear in the momentum operator, \cite{Bir-Pikus,Chuang,Sirenko} 
\begin{eqnarray} \label{Hso}
\mathbf H^{(\sigma)}= 
\begin{pmatrix}
0&-i\Delta_2 \sigma_z&i\Delta_3 \sigma_y\cr
i\Delta_2 \sigma_z&0&-i\Delta_3 \sigma_x\cr
-i\Delta_3 \sigma_y&i\Delta_3 \sigma_x&0
\end{pmatrix},
\end{eqnarray}
\newline
where $\sigma_x$, $\sigma_y$ and $\sigma_z$ are the Pauli matrices, \cite{Landavshitz-QM}
and the real functions $\Delta_2 = \Delta_2\left(\mathbf r ,z\right)$ and $\Delta_3= \Delta_3\left(\mathbf r, z\right)$ are the position-dependent parameters of the valence-band spin-orbit splitting. \cite{Bir-Pikus,Chuang,Sirenko}

The kinetic energy Hamiltonian $\mathbf H^{(k)}$ of the Eq.~(\ref{H}) is
\begin{widetext}
\begin{equation}\label{Hk}
\mathbf H^{(k)} = \begin{pmatrix}
L_1 k^2_x + M_1 k^2_y + M_2 k^2_z & N_1 k_x k_y & N_2 k_x k_z - N_3 k_x\cr
N_1 k_x k_y & M_1 k^2_x + L_1 k^2_y + M_2 k^2_z & N_2 k_y k_z - N_3 k_y\cr
N_2 k_x k_z+N_3 k_x & N_2 k_y k_z +N_3 k_y& M_3\left( k^2_x + k^2_y \right) + L_2 k^2_z
\end{pmatrix},
\end{equation}
\end{widetext}
where
\begin{equation}
\begin{split} \label{LMN_thru_A1}
&L_1 = \frac {\hbar^2}{2m_0}\left( A_2 +A_4 +A_5 \right),
\quad L_2 = \frac {\hbar^2}{2m_0}A_1, \\
&M_1 = \frac {\hbar^2}{2m_0}\left( A_2 +A_4 -A_5\right),
\quad N_1 = \frac {\hbar^2}{2m_0} 2 A_5,
\\
&M_2 = \frac {\hbar^2}{2m_0}\left( A_1+A_3 \right),
\quad N_2 = \frac {\hbar^2}{2m_0} \sqrt 2 A_6\\
&M_3 = \frac {\hbar^2}{2m_0} A_2, \quad N_3 = i \sqrt 2 A_7,
\end{split}
\end{equation}
with $A_1$, $A_2$, \ldots $A_7$ being real material parameters in conventional notations, \cite{Bir-Pikus,Chuang,Ren} $m_0$ is the free electron mass, and $\hbar k_j = -i \hbar \nabla_j$, where $j=x,y,z$, is the momentum operator. Not all the parameters are independent. The six-fold rotational symmetry of wurtzite, actually leading to isotropic symmetry of the spectrum in bulk materials in the EM approximation, \cite{Sirenko} results in the following identity: \cite{Chuang}
\begin{eqnarray}
L_1 - M_1 = N_1.
\label{L-M=N}
\end{eqnarray}

A common mistake in the literature is to treat the EM parameters as position-dependent due to variation of the chemical composition of the structure. As soon as we use the classic EM approximation, expressed as a set of second-order differential equations, we must neglect the position-dependence in the matrix-tensor of the reciprocal EMs. The values of its components that are proper for the reference material must be adopted for the whole semiconductor structure, see the details in Ref.~\onlinecite{Takhtamirov}.

Finally, the strain Hamiltonian $\mathbf H^{(\varepsilon)}$ has a structure resembling that of Eq.~(\ref{Hk}) because the tensor of the deformation potentials has the same transformation properties as the tensor of the reciprocal EMs entering Eq.~(\ref{Hk}), both being governed by the point-group  lattice symmetry, \cite{Bir-Pikus}
\begin{widetext}
\begin{equation}\label{He}
\mathbf H^{(\varepsilon)} = \begin{pmatrix}
l_1 \varepsilon_{xx} + m_1 \varepsilon_{yy} + m_2 \varepsilon_{zz} &
n_1 \varepsilon_{xy} & n_2 \varepsilon_{xz}\cr
n_1 \varepsilon_{xy} &
m_1 \varepsilon_{xx} + l_1 \varepsilon_{yy} + m_2 \varepsilon_{zz} &
n_2 \varepsilon_{yz}\cr
n_2 \varepsilon_{xz} & n_2 \varepsilon_{yz} &
m_3\left( \varepsilon_{xx} + \varepsilon_{yy}\right) + l_2 \varepsilon_{zz}
\end{pmatrix}.
\end{equation}
\end{widetext}
Here $\boldsymbol {\varepsilon} = \boldsymbol {\varepsilon} \left(\mathbf r, z\right)$ is the strain tensor, \cite{Landavshitz} and the real parameters $l_1$, $l_2$, $m_1$, $m_2$, $n_1$, and $n_2$ are expressed through the conventional \cite{Bir-Pikus,Chuang,Ren} components $D_1$, $D_2$, \ldots $D_6$ of the tensor of the deformation potentials as follows:
\begin{equation}
\begin{split} \label{lmn_thru_D}
&l_1 = D_2 +D_4 +D_5, \quad m_1 = D_2 +D_4 -D_5,\\
&n_1 = 2 D_5,\quad l_2 = D_1, \quad m_2 = D_1+D_3,\\
&n_2 = \sqrt 2 D_6, \quad m_3 = D_2.
\end{split}
\end{equation}
Analogously to the identity of Eq.~(\ref{L-M=N}), the following important relation takes place:
\begin{equation}\label{l-m=n}
l_1 - m_1 = n_1,
\end{equation}
which will secure a cylindrically symmetric strain Hamiltonian for cylindrically symmetric strains.

The deformation potentials can be treated as position-dependent because strain directly modifies the potential energy of an electron, which is a dominant effect in view of the estimation of the terms entering the Hamiltonian. \cite{Takhtamirov_sst} On the other hand, for heterostructures composed of related materials the electron band-structure parameters should have a weak position dependence, which is a requirement for the applicability of the EM approximation. The deformation potentials are also expected to have a weak position dependence in such systems, so that the full set of material parameters of the reference semiconductor, including the deformation potentials, may be used for the whole structure.

For cylindrically symmetric structures, the Hamiltonians $\mathbf H^{(0)}$ and $\mathbf H^{(\sigma)}$ along with strain $\boldsymbol {\varepsilon}$ depend on the absolute value of the vector $\mathbf r$ only. The symmetry is not yet visible in the kinetic energy Hamiltonian $\mathbf H^{(k)}$ and the strain Hamiltonian $\mathbf H^{(\varepsilon)}$.

\subsection{Hamiltonian for wurtzite in cylindrical coordinates}\label{II_wurtzite_cyl}

To transform Eq.~(\ref{H}) to cylindrical coordinates $\left(r, \phi, z\right)$, with $x = r \cos \phi$ and $y = r \sin \phi$, we first introduce new basis functions $ \widetilde {\mathbf u} = \left( u_r\ u_\phi\ u_z \right)$, where 
\begin{equation}
u_r = \frac { x u_x + yu_y } {\sqrt{x^2 + y^2}}, \quad u_\phi = \frac{ x u_y - yu_x} {\sqrt{x^2 + y^2}}.
\end{equation}
It is easily verified that they are invariant under rotation of the Cartesian coordinate system around the $z$-axis, recalling that $u_x$ and $u_y$ transform as the coordinates $x$ and $y$, respectively. In alternative notations, clarifying the mathematical sense of the new basis, the functions $ \widetilde {\mathbf u}$ are related to $\mathbf u$ as follows: $\mathbf u =  \widetilde{\mathbf u} \mathbf S$, where
\begin{equation}
\label{S}
\mathbf S =
\begin{pmatrix}
\cos \phi & \sin \phi & 0\cr
-\sin \phi & \cos \phi & 0\cr
0& 0 & 1
\end{pmatrix},
\end{equation}
which is a standard unitary matrix used to express Cartesian 3D unit vectors in terms of their cylindrical components. \cite{Korn} Note that $u_r$ and $u_\phi$ are not Bloch functions, in particular they do not possess the periodicity properties of the functions $u_x$ and $u_y$.

The total wave function, see Eq.~(\ref{psi}), is
\begin{equation} \label{psi1}
\Psi =  \widetilde {\mathbf u} \mathbf S \mathbf F \equiv   \widetilde {\mathbf u}  \widetilde {\mathbf F},
\end{equation}
where we have introduced the modified EFs $\widetilde {\mathbf F}$. Equation~(\ref{H}) now reads
\begin{equation} \label{tildeH}
 \widetilde {\mathbf H}  \widetilde {\mathbf F} = E   \widetilde {\mathbf F},
\end{equation}
where $ \widetilde {\mathbf H} = \mathbf S \mathbf H \mathbf S ^{-1}$.

Our unitary transformation does not change the potential energy Hamiltonian,
\begin{equation}\label{tildeH0}
\widetilde {\mathbf H}^{(0)} = \mathbf S \mathbf H^{(0)} \mathbf S ^{-1} = \mathbf H^{(0)}.
\end{equation}

To transform the kinetic energy Hamiltonian, presented by Eq.~(\ref{Hk}), we use Eq.~(\ref{L-M=N}) and the identities
\begin{equation} \label{nabla_j}
\begin{split}
& \nabla_x = \cos \phi \, \nabla _r
- \frac {\sin \phi } r \, \nabla _\phi,\\
&\nabla_y = \sin \phi \, \nabla _r
+ \frac {\cos \phi } r \, \nabla_ \phi,
\end{split}
\end{equation}
where $\nabla_r = \partial /\partial r$ and $\nabla_\phi = \partial /\partial \phi$.
Let us follow some algebra necessary to obtain the elements of the matrix Hamiltonian $ \widetilde {\mathbf H}^{(k)} = \mathbf S \mathbf H^{(k)} \mathbf S ^{-1}$. For example,
\begin{widetext}
\begin{equation}
\begin{split}
 \widetilde {H}^{(k)}_{11} = &\cos \phi \, H^{(k)}_{11} \cos \phi + \cos \phi\, H^{(k)}_{12}\sin \phi+ \sin \phi\,H^{(k)}_{21} \cos \phi + \sin \phi\,H^{(k)}_{22}\sin \phi\\
=& -L_1 \left( \cos \phi\,\nabla_x + \sin \phi\,\nabla_y\right)
\left( \nabla_x \cos \phi + \nabla_y \sin \phi \right)
-M_1 \left( \cos \phi\,\nabla_y - \sin \phi\,\nabla_x\right)
\left( \nabla_y \cos \phi - \nabla_x \sin \phi \right)
- M_2 \nabla_z^2\\
=&-L_1 \left( \nabla_r^2 + \nabla_r \frac 1 r \right) - M_1 \frac 1 {r^2} \nabla_\phi^2
-M_2 \nabla_z^2,
\end{split}
\end{equation}
\begin{equation}
\begin{split}
 \widetilde {H}^{(k)}_{12} = &-\cos \phi \, H^{(k)}_{11} \sin \phi + \cos \phi\, H^{(k)}_{12}\cos \phi - \sin \phi\,H^{(k)}_{21} \sin \phi + \sin \phi\,H^{(k)}_{22}\cos \phi\\
=& -L_1 \left( \cos \phi\,\nabla_x + \sin \phi\,\nabla_y\right)
\left( \nabla_y \cos \phi - \nabla_x \sin \phi \right)
-M_1 \left( \sin \phi\,\nabla_x - \cos \phi\,\nabla_y\right)
\left( \nabla_x \cos \phi + \nabla_y \sin \phi \right)\\
=&-L_1 \nabla_r \frac 1 r \nabla_ \phi + M_1 \left( \frac 1 r \nabla_r \nabla_\phi + \frac 1 {r^2}\nabla_\phi \right).
\end{split}
\end{equation}
The rest elements are obtained analogously. The full matrix of the kinetic energy Hamiltonian $ \widetilde {\mathbf H}^{(k)}$ is
\begin{equation} \label{tildeHk}
 \widetilde {\mathbf H}^{(k)} = -
\begin{pmatrix}
L_1 \left[ \nabla_r^2 + \nabla_r \frac 1 r \right] + M_1 \frac {\nabla_\phi^2} {r^2}
+ M_2 \nabla_z^2 &
L_1 \nabla_r \frac {\nabla_ \phi} r - M_1 \left[ \frac {\nabla_\phi} r \nabla_r + \frac {\nabla_\phi} {r^2} \right] &
N_2 \nabla_r \nabla_z -i N_3 \nabla_r\\
L_1 \left[ \frac {\nabla_\phi} r \nabla_r + \frac {\nabla_\phi} {r^2} \right]-
M_1 \nabla_r \frac {\nabla_ \phi} r &
M_1 \left[ \nabla_r^2 + \nabla_r \frac 1 r \right] + L_1 \frac {\nabla_\phi^2} {r^2}
+ M_2 \nabla_z^2 &
N_2 \frac {\nabla_\phi} r \nabla_z - i N_3 \frac {\nabla_\phi} r\\
N_2 \left[ \nabla_r +\frac 1 r \right]\nabla_z + i N_3 \left[ \nabla_r +\frac 1 r \right]&
N_2 \frac {\nabla_\phi} r \nabla_z + i N_3 \frac {\nabla_\phi} r&
M_3 \left[ \nabla_r^2 + \frac 1 r \nabla_r + \frac {\nabla_\phi^2} {r^2} \right]+
L_2 \nabla_z^2
\end{pmatrix}.
\end{equation}
One should remember that $\nabla _r ^{\rm T} = -\left( \nabla_r + r^{-1}\right)$ in cylindrical coordinates and the parameter $N_3$ is imaginary to verify that the Hamiltonian of Eq.~(\ref{tildeHk}) is Hermitian.

Analogously, with the help of the identity of Eq.~(\ref{l-m=n}), the strain Hamiltonian of Eq.~(\ref{He}) transforms into the Hamiltonian $ \widetilde {\mathbf H}^{(\varepsilon)} = \mathbf S \mathbf H^{(\varepsilon)} \mathbf S ^{-1}$,
\begin{equation}\label{tildeHe}
\widetilde {\mathbf H}^{(\varepsilon)} = \begin{pmatrix}
l_1 \varepsilon_{rr} + m_1 \varepsilon_{\phi\phi} + m_2 \varepsilon_{zz} &
n_1 \varepsilon_{r\phi} & n_2 \varepsilon_{rz}\cr
n_1 \varepsilon_{r\phi} &
m_1 \varepsilon_{rr} + l_1 \varepsilon_{\phi\phi} + m_2 \varepsilon_{zz} &
n_2 \varepsilon_{\phi z}\cr
n_2 \varepsilon_{rz} & n_2 \varepsilon_{\phi z} &
m_3\left( \varepsilon_{rr} + \varepsilon_{\phi\phi}\right) + l_2 \varepsilon_{zz}
\end{pmatrix},
\end{equation}
\end{widetext}
where $\boldsymbol {\varepsilon} = \boldsymbol {\varepsilon} \left(r, \phi, z \right)$ is now the strain tensor in cylindrical coordinates. \cite{Landavshitz} We have also used the following identities accompanying the coordinate change:
\begin{equation}
\begin{split}
&\varepsilon_{xx}= \varepsilon_{rr} \cos^2 \phi + \varepsilon_{\phi\phi}\sin^2 \phi - \varepsilon_{r \phi} \sin 2\phi,\cr
&\varepsilon_{yy}= \varepsilon_{rr} \sin^2 \phi + \varepsilon_{\phi\phi}\cos^2 \phi + \varepsilon_{r \phi} \sin 2\phi,\cr
&\varepsilon_{xy}= \left( \varepsilon_{rr} - \varepsilon_{\phi\phi} \right) \sin \phi \cos \phi + \varepsilon_{r \phi} \cos 2 \phi,\cr
&\varepsilon_{xz}= \varepsilon_{r z} \cos \phi - \varepsilon_{\phi z} \sin \phi ,\cr
&\varepsilon_{yz}= \varepsilon_{r z} \sin \phi + \varepsilon_{\phi z} \cos \phi ,
\end{split}
\end{equation}
which are in effect for any second-rank tensor. For cylindrically symmetric systems, the components of the strain tensor, expressed through the displacement vector ${\mathbf v} = \left( v_r, v_\phi , v_z\right)$, for which $v_\phi =0$ and $\partial {\mathbf v}/\partial \phi =0$, are \cite{Landavshitz}
\begin{equation}
\begin{split}\label{strain_cyl_sym}
&\varepsilon_{rr}= \frac {\partial v_r}{\partial r}, \quad
\varepsilon_{\phi \phi}= \frac {v_r} r, \quad
\varepsilon_{zz}= \frac {\partial v_z}{\partial z},\\
&\varepsilon_{r z}= \frac 1 2 \left( \frac {\partial v_r}{\partial z}+ \frac {\partial v_z}{\partial r} \right), \quad \varepsilon_{r \phi } = 
\varepsilon_{\phi z}= 0.
\end{split}
\end{equation}
The strain tensor can be found by solving the elasticity theory problem \cite{Andreev} in particular in cylindrical coordi\-nates. \cite{Barettin}

The kinetic energy and strain Hamiltonians $\widetilde {\mathbf H}^{(k)}$ and $\widetilde {\mathbf H}^{(\varepsilon)}$ gain the desired cylindrically symmetric form. But the spin-orbit interaction Hamiltonian of Eq.~(\ref{Hso}) transforms into the Hamiltonian $ \widetilde {\mathbf H}^{(\sigma)} = \mathbf S \mathbf H^{(\sigma)} \mathbf S ^{-1}$,
\begin{equation}
\widetilde {\mathbf H}^{(\sigma)}= 
\begin{pmatrix}
0&-i\Delta_2 \sigma_z&i\Delta_3 \sigma_\phi\cr
i\Delta_2 \sigma_z&0&-i\Delta_3 \sigma_r\cr
-i\Delta_3 \sigma_\phi&i\Delta_3 \sigma_r&0
\end{pmatrix},\label{tildeHso}
\end{equation}
where $\sigma_r = \sigma_x \exp \left( i \phi \sigma_z \right)$ and $\sigma_\phi = \sigma_y \exp \left( i \phi \sigma_z \right)$.
This Hamiltonian depends on $\phi$ and does not commute with the operator $-i\nabla_\phi$. To avoid this dependence, we note that
\begin{equation}
\mathrm e ^{ i \frac \phi 2 \sigma_z }\sigma_r
\mathrm e ^{ -i \frac \phi 2 \sigma_z } = \sigma_x,\quad
\mathrm e ^{ i \frac \phi 2 \sigma_z }\sigma_\phi
\mathrm e ^{ -i \frac \phi 2 \sigma_z } = \sigma_y,
\end{equation}
and construct another unitary transformation of the EFs,
\begin{equation} \label{tildeS}
\widetilde {\mathbf S} \widetilde {\mathbf F} = \widetilde {\widetilde {\mathbf F}},
\end{equation}
where
\begin{equation}\label{tildeSexplicit}
\widetilde {\mathbf S} = 
\begin{pmatrix}
\mathrm e ^{ i \frac \phi 2 \sigma_z }&0&0\cr
0&\mathrm e ^{ i \frac \phi 2 \sigma_z }&0\cr
0&0&\mathrm e ^{ i \frac \phi 2 \sigma_z }
\end{pmatrix}.
\end{equation}
After this unitary transformation, Eq.~(\ref{tildeH}) reads
\begin{equation} \label{tildetildeH}
 \widetilde {\widetilde {\mathbf H}}  \widetilde {\widetilde {\mathbf F}} = E   \widetilde {\widetilde {\mathbf F}},
\end{equation}
where $ \widetilde {\widetilde {\mathbf H}} = \widetilde {\mathbf S} \widetilde {\mathbf H} \widetilde {\mathbf S} ^{-1}$.
The spin-orbit interaction Hamiltonian of Eq.~(\ref{tildeHso}) recovers its initial form given by Eq.~(\ref{Hso}),
\begin{equation} \label{tildetildeHso}
\widetilde {\widetilde {\mathbf H}}\,\!^{(\sigma)} = \widetilde {\mathbf S} \widetilde {\mathbf H}^{(\sigma)} \widetilde {\mathbf S} ^{-1} = {\mathbf H}^{(\sigma)}.
\end{equation}
The potential energy Hamiltonian of Eq.~(\ref{tildeH0}) and the strain Hamiltonian of Eq.~(\ref{tildeHe}) remain unchanged,
\begin{equation} \label{tildetildeH0+He}
\widetilde {\widetilde {\mathbf H}}\,\!^{(0)} +
\widetilde {\widetilde {\mathbf H}}\,\!^{(\varepsilon)}
 = \widetilde {\mathbf S} \left( 
{\mathbf H}^{(0)} + \widetilde {\mathbf H}^{(\varepsilon)} \right)
\widetilde {\mathbf S} ^{-1} =
{\mathbf H}^{(0)} + \widetilde {\mathbf H}^{(\varepsilon)}.
\end{equation}
It follows from the identity
\begin{equation}
\mathrm e ^{ i \frac \phi 2 \sigma_z } \nabla _\phi
\mathrm e ^{ -i \frac \phi 2 \sigma_z } = \nabla _\phi - \frac i 2 \sigma_z
\end{equation}
that the kinetic energy Hamiltonian $\widetilde {\widetilde {\mathbf H}}\,\!^{(k)} = \widetilde {\mathbf S} \widetilde {\mathbf H}^{(k)} \widetilde {\mathbf S} ^{-1}$ still has the form given by Eq.~(\ref{tildeHk}) where the operator $\nabla_\phi$ is replaced by the matrix operator
\begin{equation}\label{nabla_matrix}
\widehat {\boldsymbol \nabla}_\phi =
\begin{pmatrix}
\nabla_\phi - \frac i 2&0\cr
0&\nabla_\phi + \frac i 2
\end{pmatrix};
\end{equation}
we express this by using the following notations:
\begin{equation}\label{nabla-bold}
\widetilde {\widetilde {\mathbf H}}\,\!^{(k)} = \widetilde {\mathbf H} ^{(k)} ( \nabla _\phi \rightarrow \widehat {\boldsymbol \nabla}_\phi ).
\end{equation}

For cylindrically symmetric systems, the Hamiltonian $\widetilde {\widetilde {\mathbf H}}$ now commutes with the operator of the $z$-projection of angular momentum $j_z = -i\nabla_\phi$ [in units of $\hbar$]. Note that we could bypass the unitary transformation of Eq.~(\ref{tildeS}), but use the fact that the reciprocal transformation of the operator $j_z$,
\begin{equation}
J_z =  \mathrm e ^{ -i \frac \phi 2 \sigma_z } j_z \, \mathrm e ^{ i \frac \phi 2 \sigma_z } = \frac {\nabla_\phi} i + \frac 1 2 \sigma_z,
\end{equation}
produces the operator of the $z$-projection of {\em total} angular momentum $J_z$ that commutes with the Hamiltonian $\widetilde {\mathbf H}$. The commuting operators $\widetilde {\widetilde {\mathbf H}}$ and $j_z$ have common eigenfunctions, \cite{Landavshitz-QM} so that the EFs $\widetilde {\widetilde {\mathbf F}} = \widetilde {\widetilde {\mathbf F}} \left( r, \phi,z \right)$ can be chosen in the following form:
\begin{equation} \label{tildetildeF}
\widetilde {\widetilde {\mathbf F}} = \frac {{\mathrm e}^{im\phi}}{\sqrt {2\pi}}\, \mathbf f \left(r, z \right),
\end{equation}
where $m$, which is an eigenvalue of the operator of projection of the electron's total angular momentum on the symmetry axis, should be half-integer, $m = \pm 1/2, \pm 3/2, \dots$ The functions
$\mathbf f  = \mathbf f \left(r, z \right) = \left( \mathbf f_1 \ \mathbf f_2 \ \mathbf f_3 \right)^{\mathrm T}$,
\begin{equation}
\mathbf f_j = \left( f^{(u)}_j \ f^{(d)}_j \right)^{\mathrm T}, \quad j=1,2,3,
\end{equation}
satisfy the equation
\begin{equation}\label{Hf_Ef}
\left( \mathbf H^{(0)} + {\mathbf H}^{(\sigma)} + \widetilde {\mathbf H}^{(\varepsilon)} +\widetilde {\widetilde {\mathbf H}}\,\!^{(k)}_m \right) \mathbf f = E \mathbf f,
\end{equation}
where $\mathbf H^{(0)}$, ${\mathbf H}^{(\sigma)}$ and $\widetilde {\mathbf H}^{(\varepsilon)}$ are given by Eqs.~(\ref{H0}), (\ref{Hso}) and (\ref{tildeHe}), respectively, and the stain Hamiltonian being supplemented by Eq.~(\ref{strain_cyl_sym}). Having $m$ as a parameter entering the matrix
\begin{equation}
{\mathbf m} = 
\begin{pmatrix}
m - \frac 1 2 &0\cr
0& m +\frac 1 2
\end{pmatrix},
\end{equation}
the kinetic energy Hamiltonian is
\begin{widetext}
\begin{equation} \label{tildeHkm}
\widetilde {\widetilde {\mathbf H}}\,\!^{(k)}_m = -
\begin{pmatrix}
L_1 \left[ \nabla_r^2 + \nabla_r \frac 1 r \right] - M_1 \frac {{\mathbf m}^2} {r^2}
+ M_2 \nabla_z^2 &
i L_1 \nabla_r \frac {{\mathbf m}} r - i M_1 \left[ \frac {{\mathbf m}} r \nabla_r + \frac {{\mathbf m}} {r^2} \right] &
N_2 \nabla_r \nabla_z - i N_3 \nabla_r\\
i L_1 \left[ \frac {{\mathbf m}} r \nabla_r + \frac {{\mathbf m}} {r^2} \right]-
i M_1 \nabla_r \frac {{\mathbf m}} r &
M_1 \left[ \nabla_r^2 + \nabla_r \frac 1 r \right] - L_1 \frac {{\mathbf m}^2} {r^2}
+ M_2 \nabla_z^2 &
i N_2 \frac {{\mathbf m}} r \nabla_z + N_3 \frac {{\mathbf m}} r\\
N_2 \left[ \nabla_r +\frac 1 r \right]\nabla_z + i N_3 \left[ \nabla_r +\frac 1 r \right]&
i N_2 \frac {{\mathbf m}} r \nabla_z - N_3 \frac {{\mathbf m}} r&
M_3 \left[ \nabla_r^2 + \frac 1 r \nabla_r - \frac {{\mathbf m}^2} {r^2} \right]+
L_2 \nabla_z^2
\end{pmatrix}.
\end{equation}
\end{widetext}

The total wave function $\Psi$ and the EFs $\widetilde {\widetilde {\mathbf F}}$ are related to each other through the conventional basis functions $\mathbf u$ or the modified ones $\widetilde {\mathbf u}$ as follows:
\begin{equation} \label{psi2}
\Psi =  {\mathbf u} {\mathbf S}^{-1} \widetilde {\mathbf S}^{-1} \widetilde {\widetilde {\mathbf F}} = \widetilde {\mathbf u} \widetilde {\mathbf S}^{-1} \widetilde {\widetilde {\mathbf F}},
\end{equation}
where the unitary matrices ${\mathbf S}$ and $\widetilde {\mathbf S}$ are given by Eqs.~(\ref{S}) and (\ref{tildeSexplicit}), respectively. For the basis functions $u_r$ and $u_\phi$ are not periodic, it may be convenient to retain the old basis function ${\mathbf u}$, which are periodic and orthonormal, recovering the old EFs $\mathbf F$,
\begin{equation} \label{old_F}
\mathbf F = {\mathbf S}^{-1} \widetilde {\mathbf S}^{-1} \widetilde {\widetilde {\mathbf F}},
\end{equation}
as soon as the system of the EM equation has been solved and the EFs $\widetilde {\widetilde {\mathbf F}}$ have been found.

An external magnetic field applied along the symmetry axis does not break the cylindrical symmetry of the system present. The corresponding formalism is given in Appendix~\ref{II_mf}.

\subsection{Boundary conditions on symmetry axis}\label{II_bc}

A transformation of the EM equations from Cartesian coordinates to spherical or cylindrical ones calls for BCs to be imposed on the EFs at the origin or on the symmetry axis, respectively. In the single-band case, such BCs are either zero EF or zero slope of the EF, depending on eigenvalues of the angular momentum. \cite{Messiah,Landavshitz-QM} We will now determine the BCs for the functions $\mathbf f \left( r,z\right)$, which satisfy Eq.~(\ref{Hf_Ef}), on the symmetry axis $r=0$. This will complete our formulation of the valence-band Hamiltonian in cylindrical coordinates. To fully specify the eigenvalue problem, one should also set the BCs at infinity, e.g., $\mathbf f \left( r,z\right) \vert _{r \rightarrow \infty} \rightarrow 0$ for discrete spectrum.

The usual approach to this problem, applied for the single-band equation \cite{Messiah,Landavshitz-QM} and related to the Frobenius method of solving second-order ordinary differential equations, \cite{Frobenius} is a very complicated way for our multiband case. Instead, we obtain the BCs from the continuity of the functions $\mathbf F$, which satisfy Eq.~(\ref{H}), and their first partial derivatives $\nabla_j \mathbf F$ for $j=x,y$. Expanding Eqs.~(\ref{tildetildeF}) and (\ref{old_F}), we have the following relations:
\begin{equation}\label{u_components}
\begin{split}
&\sqrt {2\pi} F^{(u)}_x = f^{(u)}_1 \mathrm e^{i\mu \phi} \cos \phi - f^{(u)}_2 \mathrm e^{i\mu \phi} \sin \phi,\\
&\sqrt {2\pi} F^{(u)}_y = f^{(u)}_1 \mathrm e^{i\mu \phi} \sin \phi + f^{(u)}_2 \mathrm e^{i\mu \phi} \cos \phi,\\
& \sqrt {2\pi}  F^{(u)}_z = f^{(u)}_3 \mathrm e^{i\mu \phi},
\end{split}
\end{equation}
for the upper elements of the spinor components, where the integer $\mu = m-1/2$, and similar expressions for the down elements, where the EF index $(u)$ is substituted by $(d)$ and the integer $\nu = m + 1/2$ supersedes $\mu$.

The continuity of the functions $\mathbf F$ in particular on the symmetry axis implies that 
\begin{equation}\label{continuity_F}
\mathbf F \left( \mathbf r ,z \right)\vert_{x=r\cos \phi , y=r \sin \phi} \rightarrow
\mathbf F \left( \mathbf r ,z \right)\vert_{x=r, y=0},
\end{equation}
when $r \rightarrow 0$ for any angle $\phi$. For the function $f^{(u)}_3$, using Eqs.~(\ref{u_components}) and (\ref{continuity_F}), we have the following:
\begin{equation}
\mathrm e^{i\mu \phi} f^{(u)}_3 \left( r , z\right) \rightarrow
f^{(u)}_3 \left( r , z\right), \quad r  \rightarrow 0, \ \forall \ \phi,
\end{equation}
which is an identity if $\mu = 0$, but definitely
\begin{equation}\label{1st_BC_3u}
f^{(u)}_3 \vert _{r = 0} = 0, \quad \mu \ne 0.
\end{equation}
For the down element of the spinor we obtain analogously
\begin{equation}\label{1st_BC_3d}
f^{(d)}_3 \vert _{r = 0} = 0, \quad \nu \ne 0.
\end{equation}
From the continuity condition of Eq.~(\ref{continuity_F}), using the first identity in Eq.~(\ref{u_components}), we have for any $\phi$
\begin{equation}
f^{(u)}_1 \mathrm e^{i\mu \phi} \cos \phi - f^{(u)}_2 \mathrm e^{i\mu \phi} \sin \phi \rightarrow f^{(u)}_1, \quad r \rightarrow 0,
\end{equation}
which results in the following:
\begin{equation}\label{1st_BC_12u}
\begin{split}
&f^{(u)}_1 \vert _{r = 0} = f^{(u)}_2 \vert _{r = 0} =0, \quad \vert \mu \vert \ne 1,\\
&f^{(u)}_2 \vert _{r = 0} = i \mu f^{(u)}_1 \vert _{r = 0}, \quad \vert \mu \vert = 1.
\end{split}
\end{equation}
For the down elements of the spinor it holds analogously that
\begin{equation}\label{1st_BC_12d}
\begin{split}
&f^{(d)}_1 \vert _{r = 0} = f^{(d)}_2 \vert _{r = 0} =0, \quad \vert \nu \vert \ne 1,\\
&f^{(d)}_2 \vert _{r = 0} = i \nu f^{(d)}_1 \vert _{r = 0}, \quad \vert \nu \vert = 1.
\end{split}
\end{equation}
We will have the same results if we analyze the continuity of the functions $F^{(u)}_y$ and $F^{(d)}_y$.

To obtain the BCs imposed on the derivatives of the EFs, we use the following continuity conditions for $r \rightarrow 0$ and any $\phi$:
\begin{equation}\label{continuity_F_prime}
\nabla_j \mathbf F \left( \mathbf r ,z \right)\vert_{x=r\cos \phi ,  y=r \sin \phi}
\rightarrow  \nabla_j \mathbf F \left( \mathbf r ,z \right)\vert_{x=r, y=0},
\end{equation}
where $j=x,y$. For the element $f^{(u)}_3$, using its representation                                                                            through $F^{(u)}_z$ in Eq.~(\ref{u_components}), as well as the conditions of Eq.~(\ref{continuity_F_prime}) and the identities of Eq.~(\ref{nabla_j}), we have in particular the following:
\begin{equation}\label{f_3_prime}
\frac {\partial f^{(u)}_3}{\partial r} \mathrm e^{i\mu \phi} \cos \phi - i \mu f^{(u)}_3 \frac {\sin \phi}{r} \mathrm e^{i\mu \phi} \rightarrow \frac {\partial f^{(u)}_3}{\partial r},
\end{equation}
when $r \rightarrow 0$ for any $\phi$, from which we conclude that
\begin{equation} \label{2nd_BC_3u}
\nabla _r f^{(u)}_3 \vert_{r = 0} = 0, \quad \mu =0.
\end{equation}
Analogously, for the down element,
\begin{equation} \label{2nd_BC_3d}
\nabla _r f^{(d)}_3 \vert_{r = 0} = 0, \quad \nu =0.
\end{equation}
For the elements $f^{(u)}_1$ and $f^{(u)}_2$, we use the continuity conditions of Eq.~(\ref{continuity_F_prime}), the first and the second identities of Eq.~(\ref{u_components}) along with the identities of Eq.~(\ref{nabla_j}). We obtain in particular that for any $\phi$ and $r \rightarrow 0$
\begin{equation}\label{f_12_prime}
\begin{split}
&\frac {\partial f^{(u)}_1}{\partial r} \mathrm e^{i\mu \phi} \cos^2 \phi
- \frac {\partial f^{(u)}_2}{\partial r} \mathrm e^{i\mu \phi} \sin \phi \cos \phi \\
&+ \frac {f^{(u)}_1}{r} \mathrm e^{i\mu \phi} \left( \sin^2 \phi - i \mu \sin \phi \cos \phi \right) \\
& + \frac {f^{(u)}_2}{r} \mathrm e^{i\mu \phi} \left( i \mu \sin^2 \phi + \sin \phi \cos \phi \right)
\rightarrow \frac {\partial f^{(u)}_1}{\partial r}.
\end{split}
\end{equation}
For $\vert \mu \vert = 1$, using Eqs.~(\ref{1st_BC_12u}) and (\ref{f_12_prime}), we have
\begin{equation}\label{f_12_prime_simple}
\frac {\partial f^{(u)}_1}{\partial r} \mathrm e^{i\mu \phi} \cos^2 \phi
- \frac {\partial f^{(u)}_2}{\partial r} \mathrm e^{i\mu \phi} \sin \phi \cos \phi
\rightarrow \frac {\partial f^{(u)}_1}{\partial r},
\end{equation}
which is satisfied for any $\phi$ and $r \rightarrow 0$ if
\begin{equation}\label{2nd_BC_12u}
\nabla _r f^{(u)}_1 \vert_{r = 0} = \nabla _r f^{(u)}_2 \vert_{r = 0} = 0, \quad \vert \mu \vert =1.
\end{equation}
For the down elements we have
\begin{equation}\label{2nd_BC_12d}
\nabla _r f^{(d)}_1 \vert_{r = 0} = \nabla _r f^{(d)}_2 \vert_{r = 0} = 0, \quad \vert \nu \vert =1.
\end{equation}
\begin{table}[bth]
\squeezetable
\caption{Boundary conditions imposed on $\mathbf f \left( r,z\right)$ at $r =0$ for different eigenvalues $m$ of the operator of projection of total angular momentum on the symmetry axis.}
\begin{ruledtabular}
\begin{tabular}{lccccc}
$m$ & 1/2 & --1/2 & 3/2 & --3/2 & other half-integers\\
\hline
$f^{(u)}_1$           &  0 &  ---              & ---          & 0   & 0\\
$f^{(u)}_2$           &  0 &  $-if^{(u)}_1$    & $if^{(u)}_1$ & 0   & 0\\
$f^{(u)}_3$           &  --- &  0              & 0            & 0   & 0\\
$\nabla _r f^{(u)}_1$ &  --- &  0              & 0            & --- & ---\\
$\nabla _r f^{(u)}_2$ &  --- & (0)            & (0)            & --- & ---\\
$\nabla _r f^{(u)}_3$ &  0   &  ---            & ---          & --- & ---\\
$f^{(d)}_1$           &  ---          &  0   & 0   & ---           & 0\\
$f^{(d)}_2$           &  $if^{(d)}_1$ &  0   & 0   & $-if^{(d)}_1$ & 0\\
$f^{(d)}_3$           &  0            &  --- & 0   & 0             & 0\\
$\nabla _r f^{(d)}_1$ &  0            &  --- & --- & 0             & ---\\
$\nabla _r f^{(d)}_2$ & (0)            &  --- & --- & (0)          & ---\\
$\nabla _r f^{(d)}_3$ &  ---          &  0   & --- & ---           & ---
\end{tabular}
\end{ruledtabular}
\label{BC6x6}
\end{table}

The conditions of Eqs.~(\ref{1st_BC_3u}), (\ref{1st_BC_3d}), (\ref{1st_BC_12u}), (\ref{1st_BC_12d}), (\ref{2nd_BC_3u}), (\ref{2nd_BC_3d}), (\ref{2nd_BC_12u}), and (\ref{2nd_BC_12d}) constitute a complete set of the BCs that must be imposed on the functions $\mathbf f \left( r,z\right)$ on the symmetry axis. Not all the BCs are independent though. For example, we have as many as four BCs at $r=0$ for the upper spinor elements for $m=-1/2$,
\begin{equation}\label{bc_f2f1f3}
f^{(u)}_2 = -if^{(u)}_1, \quad f^{(u)}_3=0,
\end{equation}
and
\begin{equation}\label{bc_nablaf1nablaf2}
\nabla _r f^{(u)}_1=0, \quad \nabla _r f^{(u)}_2=0,
\end{equation}
but only three BCs should be specified. One of the conditions must follow from Eq.~(\ref{Hf_Ef}) if we take into account other three. Indeed, if we multiply the second equation of the system of Eq.~(\ref{Hf_Ef}) by $i$, extract the resulting equation from the first one, and consider the formal limit $r\rightarrow 0$, taking into account the conditions in Eq.~(\ref{bc_f2f1f3}), we obtain the following result for the upper spinor elements:
\begin{equation}
M_1 \nabla _r f^{(u)}_1 - i L_1 \nabla _r f^{(u)}_2 \rightarrow 0,
\end{equation}
so that the second BC in Eq.~(\ref{bc_nablaf1nablaf2}) follows from the first one and vice versa. Other cases of superfluous BCs are treated analogously. We summarize the results in Table~\ref{BC6x6}, where the superfluous BCs are enclosed in parentheses.

The BCs are not changed in the presence of the external magnetic field, when the functions $\mathbf f \left( r,z\right)$ satisfy Eq.~(\ref{shroedinger_mf}), because Eqs.~(\ref{Hf_Ef}) and (\ref{shroedinger_mf}) are equivalent in the limit $r\rightarrow 0$ in the absence of the effective Pauli term of Eq.~(\ref{HB}), which is not singular.

\subsection{Zinc-blende}\label{II_zb}

Our results can easily be modified to be used for cylindrically symmetric structures composed of zinc-blende semiconductors. The basis functions $u_x$, $u_y$ and $u_z$ now belong to the representation $\Gamma_{15}$ of the space group $T_d$, so we put $U_{v6} = U_{v1}$ in the potential energy Hamiltonian of Eq.~(\ref{H0}), and $\Delta_2 = \Delta_3 = \Delta/3$ in the spin-orbit interaction Hamiltonians of Eqs.~(\ref{Hso}) and (\ref{tildeHso}), where $\Delta$ is the valence-band spin-orbit splitting in cubic materials. \cite{Luttinger}

For the parameters entering the kinetic energy Hamiltonians of Eqs.~(\ref{Hk}), (\ref{tildeHk}), (\ref{tildeHkm}), and (\ref{tildeHkmB}), we have the following identities:
\begin{equation} \label{kinetic_cubic}
L_1 = L_2, \ \ M_1 = M_2=M_3, \ \ N_1 = N_2, \ \ N_3 = 0.
\end{equation}
Three independent parameters $L_1$, $M_1$, and $N_1$ are expressed through the Luttinger parameters $\gamma_1$, $\gamma_2$, and $\gamma_3$ as follows: \cite{Luttinger2}
\begin{equation} \label{Luttinger_param}
\begin{split}
\frac {\gamma_1}{2m_0} = - \frac 1 3 \left( L_1 + 2M_1\right), &\ \ 
\frac {\gamma_2}{2m_0} = - \frac 1 6 \left( L_1 - M_1\right), \\
\frac {\gamma_3}{2m_0} &= - \frac 1 6 N_1.
\end{split}
\end{equation}
Equation~(\ref{L-M=N}), being an equivalent to $\gamma_2 = \gamma_3$, is not an identity for zinc-blende but the recipe for the spherical band approximation. \cite{Baldereschi}

Analogously, in the strain Hamiltonians of Eqs.~(\ref{He}) and (\ref{tildeHe}) we have
\begin{equation} \label{strain_cubic}
l_1 = l_2, \quad  m_1 = m_2=m_3, \quad n_1 = n_2.
\end{equation}
The conventional deformation potentials $a$, $b$, and $d$ are related to these parameters as follows: \cite{Bahder,Bahder_E}
\begin{equation} \label{deformation_param}
a = -\frac 1 3 \left( l_1 + 2m_1\right), \ \ b= \frac 1 3 \left( l_1 - m_1 \right), \ \ d = \frac 1 {\sqrt 3} n_1.
\end{equation}
\begin{table}[tb]
\squeezetable
\caption{Values of the valence-band deformation potentials, taken from Ref.~\onlinecite{Vurgaftman_old}, and parameters $\mu_\varepsilon$ and $\gamma_\varepsilon$ for some zinc-blende semiconductors.}
\begin{ruledtabular}
\begin{tabular}{lrrrrrrrrr}
 & GaAs & AlAs & InAs & GaP & AlP & InP & GaN & AlN & InN\\
\hline
$a\!$ (eV) &  --1.16 &  --2.47 & --1.00 & --1.7 & --3.0 & --0.6 & --5.2 & --3.4 & --1.5 \\
$b\!$ (eV) &  --2.0 &  --2.3 & --1.8 & --1.6 & --1.5 & --2.0 & --2.2 & --1.9 & --1.2 \\
$d\!$ (eV) &  --4.8 &  --3.4 & --3.6 & --4.6 & --4.6 & --5.0 & --3.4 & --10 & --9.3 \\
$\mu_\varepsilon$ &  --2.1 &  --0.85 & --2.0 & --1.3 & --0.73 & --4 & --0.40 & --1.2 & --2.5\\
$\gamma_\varepsilon$ &  --0.33 &  0.068 & --0.14 & --0.31 & --0.19 & --0.7 & 0.023 & --0.57 & --1.4
\end{tabular}
\end{ruledtabular}
\label{bd}
\end{table}
The zinc-blende symmetry does not guarantee the validity of Eq.~(\ref{l-m=n}), an equivalent to
\begin{equation}\label{spherical_strain}
d = \sqrt 3 \, b,
\end{equation}
which may be dubbed the spherical deformation potentials approximation by analogy with the spherical band approximation. The decision on whether Eq.~(\ref{spherical_strain}) is an acceptable approximation or not should be made for each specific case.

To judge on the applicability of the spherical band approximation, Baldereschi and Lipari have introduced the following parameters:\cite{Baldereschi}
\begin{equation}
\begin{split}
\mu = &\frac {6\gamma_3+4\gamma_2}{5\gamma_1}=\frac{3N_1+2L_1-2M_1}{5L_1+10M_1},\\
\gamma = &\frac {\gamma_3- \gamma_2}{\gamma_1}=\frac{N_1-L_1+M_1}{2L_1+4M_1}.
\end{split}
\end{equation}
The cubic contribution, which is proportional to $\gamma$, is small if $\vert \gamma \vert \ll \vert \mu \vert$. This strong inequality is satisfied for a number of semiconductors.\cite{Baldereschi} As soon as the tensor of the deformation potentials has the same transformation properties as the tensor of the reciprocal EMs, see Eqs.~(\ref{Hk}) and (\ref{He}), we introduce the following parameters:
\begin{equation}
\begin{split}
\mu_\varepsilon = &\frac{3n_1+2l_1-2m_1}{5l_1+10m_1}=-\frac{\sqrt{3}d+2b}{5a},\\
\gamma_\varepsilon = &\frac{n_1-l_1+m_1}{2l_1+4m_1} = \frac{3b-\sqrt{3}d}{6a}.
\end{split}
\end{equation}
Consequently, if $\vert \gamma_\varepsilon \vert \ll \vert \mu_\varepsilon \vert$, the spherical deformation potentials approximation is expected to be accurate. In Table~\ref{bd} we present values of the valence band deformation potentials and the parameters $\mu_\varepsilon$ and $\gamma_\varepsilon$ for some zinc-blende semiconductors. Except for AlN and InN ($\mu_\varepsilon/\gamma_\varepsilon \approx 2$), the strong inequality $\vert \gamma_\varepsilon \vert \ll \vert \mu_\varepsilon \vert$ takes place for the listed materials. The spherical deformation potentials approximation can thus be as good as the spherical band approximation provided the strain distribution, which is a solution of the elasticity theory problem, is approximated as radially symmetric.

\section{Hamiltonian for Kane model}\label{III_Kane}

\subsection{Cartesian coordinates}\label{III_cart}

The Kane model \cite{Kane} takes into account the direct mixing of the valence and conduction band states exactly while the mixing mediated by remote bands is treated to the second order of the L\"owdin perturbation scheme. \cite{Bir-Pikus,Lowdin} The model is used for narrow-bandgap systems when the energies of interest, e.g., band offsets, are comparable with the bandgap, but other bands can still be treated as remote. There exists a controversy on the form of the Kane Hamiltonian for wurtzite, \cite{Rinke,Fu} though concerning minor contributions for which the corresponding material parameters have not been experimentally established yet. Besides, the wurtzite heterostructure potential acquires a non-diagonal component because the conduction and one of the valence band edge Bloch functions belong to the same representation $\Gamma_1$ of the space group $C_{6v}$. We highlight these issues below. 

The system of equations of the Kane model is
\begin{equation}\label{Schkane}
\mathbf H ' \mathbf F ' = E\mathbf F ',
\end{equation}
where
\begin{equation}\label{Hkane}
\mathbf H ' = 
\begin{pmatrix}
H_c &\mathbf H_{cv}\cr
\mathbf H^\dag _{cv}& \mathbf H_v
\end{pmatrix}, \quad \mathbf F ' = 
\begin{pmatrix}
F_c\cr
\mathbf F_v
\end{pmatrix},
\end{equation}
and $\mathbf H^\dag _{cv}$ is the Hermitian conjugate of $\mathbf H_{cv}$. The con\-duc\-tion and the valence-band EFs are $F_c  = F_c \left( \mathbf r, z \right)$ and $\mathbf F_v = \mathbf F_v \left( \mathbf r, z\right)$, respectively, the components of the latter are $F_x$, $F_y$, and $F_z$. The total wave function is
\begin{equation} \label{psi-new}
\Psi = \sum_{j=c,x,y,z} u_j F_j = \mathbf u' \mathbf F',
\end{equation}
where we have also included the conduction band zone-center Bloch function $u_c =  u_c\left( \mathbf r, z\right)$, so that $\mathbf u' = \left(u_c\ u_x\ u_y\ u_z\right)$. The basis functions $\mathbf u'$, which can be chosen real, are spinless while each EF component $F_j$ is a spinor,
\begin{equation}
F_j =
\begin{pmatrix}
F^{(u)}_j\\
F^{(d)}_j
\end{pmatrix}, \quad j=c,x,y,z.
\end{equation}

We now proceed to the description of the matrix Hamiltonian $\mathbf H '$. The conduction-band block is
\begin{equation}\label{Hc}
\begin{split}
H_c = &A'_1 k_z^2 + A'_2\left( k_x^2 +k_y^2 \right)+ U_c\\
&+a_1 \varepsilon_{zz} + a_2 \left( \varepsilon_{xx} + \varepsilon_{yy}\right),
\end{split}
\end{equation}
where $U_c = U_c\left(\mathbf r, z\right)$ is the position-dependent edge of the conduction band $\Gamma_1$, which includes an external scalar potential present, $a_1$ and $a_2$ are the deformation potentials for the conduction band. The parameters $A'_1$ and $A'_2$ are expressed through the components $1/m_1$ and $1/m_2$ of the tensor of the reciprocal EM for the conduction band in the single-band approximation and the Kane parameters $P_1 = -i \hbar \langle u_c \vert \hbar k_z \vert u_z \rangle/m_0$ and $P_2 = -i \hbar \langle u_c \vert \hbar k_x \vert u_x \rangle /m_0$ as follows:
\begin{equation}\label{A1}
A'_1 = \frac {\hbar^2}{2m_1} - \frac{P^2_1}{E_c - E_{v1} },
\end{equation}
\begin{equation}\label{A2}
A'_2 = \frac {\hbar^2}{2m_2} - \frac{P^2_2}{E_c - E_{v6} },
\end{equation}
where $E_c$, $E_{v1}$, and $E_{v6}$ are the energies of the conduction $\Gamma_1$ and valence $\Gamma_1$ and $\Gamma_6$ states of the reference material, respectively.

The valence-band block, see Eq.~(\ref{Hkane}), is
\begin{equation}\label{Hv}
\mathbf H_v = \mathbf H^{(0)} + \mathbf H^{(\sigma)} + \mathbf H^{(\varepsilon)} +\mathbf H^{\prime (k)}, 
\end{equation}
where $\mathbf H^{(0)}$, $\mathbf H^{(\sigma)}$ and $\mathbf H^{(\varepsilon)}$ are given by Eqs.~(\ref{H0}), (\ref{Hso}) and (\ref{He}), respectively. The kinetic energy Hamiltonian $\mathbf H^{\prime (k)}$ preserves the form of Eq.~(\ref{Hk}) while some of the parameters are adjusted to exclude contributions due to the mixing with the conduction band states,
\begin{widetext}
\begin{equation}\label{Hprimek}
\mathbf H^{\prime(k)} = \begin{pmatrix}
L'_1 k^2_x + M_1 k^2_y + M_2 k^2_z & N'_1 k_x k_y & N'_2 k_x k_z - N_3 k_x\cr
N'_1 k_x k_y & M_1 k^2_x + L'_1 k^2_y + M_2 k^2_z & N'_2 k_y k_z - N_3 k_y\cr
N'_2 k_x k_z+N_3 k_x & N'_2 k_y k_z +N_3 k_y& M_3\left( k^2_x + k^2_y \right) + L'_2 k^2_z
\end{pmatrix},
\end{equation}
\end{widetext}
where
\begin{equation}\label{LNprimes1}
L'_1 = L_1 + \frac{P^2_2}{E_c - E_{v6} },
\quad L'_2 = L_2 + \frac{P^2_1}{E_c - E_{v1} },
\end{equation}
\begin{equation}\label{LNprimes2}
N'_1 = N_1 + \frac{P^2_2}{E_c - E_{v6} },
\end{equation}
\begin{equation}\label{LNprimes3}
N'_2 = N_2 + \frac{P_1 P_2 \left( 2 E_c - E_{v1} - E_{v6} \right) }{2\left( E_c - E_{v6} \right)\left( E_c - E_{v1} \right)}.
\end{equation}

The block $\mathbf H_{cv}$ of the Hamiltonian of Eq.~(\ref{Hkane}) is
\begin{equation}\label{Hcv}
\mathbf H_{cv} = \begin{pmatrix}
H_{cv1} & H_{cv2} & H_{cv3} \end{pmatrix},
\end{equation}
where
\begin{equation}\label{Hcx}
H_{cv1} = i P_2 k_x + B_2 k_x k_z + b_2 \varepsilon_{xz},
\end{equation}
\begin{equation}\label{Hcy}
H_{cv2} = iP_2 k_y + B_2 k_y k_z + b_2 \varepsilon_{yz},
\end{equation}
\begin{equation}\label{Hcz}
\begin{split}
H_{cv3} =&\ iP_1 k_z + B_1 k^2_z + B_3 \left( k^2_x + k^2_y\right)\\
&+ b_1 \varepsilon_{zz} + b_3 \left( \varepsilon_{xx} + \varepsilon_{yy} \right) + U_{cz},
\end{split}
\end{equation}
where we have neglected strain-induced $\mathbf {k \cdot p}$ terms, \cite{Bir-Pikus,Bahder} proportional to $\boldsymbol {\varepsilon}$, which are small because for typical semiconductor heterostructures $\boldsymbol {\varepsilon} \sim 0.01$. The parameters $b_1=\langle u_c \vert D^{zz} \vert u_z \rangle$, $b_2=\langle u_c \vert D^{xz} \vert u_x \rangle$, and $b_3=\langle u_c \vert D^{xx} \vert u_z \rangle$ are the interband matrix elements of the deformation potentials tensor $\mathbf D$. \cite{Bir-Pikus,Bahder} The parameters $B_1$, $B_2$ and $B_3$ are
\begin{equation}\label{B1}
B_1= \frac {\hbar^2}{2m_0^2}\sum_\gamma \left( \frac 1 {E_{v1} - E_\gamma} + \frac 1 {E_c - E_\gamma} \right) p^{(z)}_{c \gamma}p^{(z)}_{\gamma z},
\end{equation}
\begin{equation}\label{B2}
\begin{split}
B_2= \frac {\hbar^2}{2m_0^2}\sum_\gamma &\left( \frac 1 {E_{v1} - E_\gamma} + \frac 1 {E_c - E_\gamma} \right)\\
&\times \left( p^{(x)}_{c \gamma}p^{(z)}_{\gamma z} + p^{(z)}_{c \gamma}p^{(x)}_{\gamma z}\right),
\end{split}
\end{equation}
\begin{equation}\label{B3}
B_3= \frac {\hbar^2}{2m_0^2}\sum_\gamma \left( \frac 1 {E_{v1} - E_\gamma} + \frac 1 {E_c - E_\gamma} \right) p^{(x)}_{c \gamma}p^{(x)}_{\gamma z}.
\end{equation}
We can now mention that the Kane Hamiltonian in Eq.~(A4) of Ref.~\onlinecite{Rinke} is not applicable to wurtzite because the terms providing the interband mixing mediated by remote bands, which are proportional to the parameters $B_1$ and $B_2$ there, are not represented correctly; the Kane Hamiltonian in Eq.~(4) of Ref.~\onlinecite{Fu} lacks the term proportional to $B_1$ in our Eq.~(\ref{Hcz}). Finalizing with the description of the Kane Hamiltonian, in Eq.~(\ref{Hcz}) we have also included the interband potential $U_{cz} = U_{cz} \left( \mathbf r , z\right)$, which exists due to the difference in the Bloch functions for the semiconductors composing the heterostructure. We set $U_{cz} =0$ for the space region of the potential well semiconductor, which has the periodic lattice potential $U_1 = U_1\left( \mathbf r ,z \right)$, while for the space region of the barrier material with the lattice potential $U_2 = U_2\left( \mathbf r ,z \right)$ we have $U_{cz} = \langle u_c \vert U_2 - U_1 \vert u_z \rangle$. Based on bulk parameters of the pair GaN/AlN, \cite{Vurgaftman_old} an estimation, details of which will be presented elsewhere, gives the value of the matrix element $\langle u_c \vert U_2 - U_1 \vert u_z \rangle$ as small as $60$~meV. It has the same order of magnitude as the parameters of the crystal splitting $\Delta_1=E_{v6}-E_{v1}$ and spin-orbit interaction $\Delta_2$ and $\Delta_3$ for wurtzite GaN and AlN. 

If we adopt the cubic approximation, \cite{Bir-Pikus,Chuang,Sirenko} which imposes the zinc-blende crystal symmetry on potentials of the wurtzite semiconductors, all parameters corresponding to the controversial terms as well as the interband offset will turn to zero,
\begin{equation}\label{nuli}
B_1 =B_2= B_3 = b_1 = b_2 = b_3 = U_{cz} = 0.
\end{equation}
Similar contributions in zinc-blende materials, \cite{Bahder} which disappear for wurtzite, can hence be treated as small on reciprocal grounds, rather than by invoking an affinity of the zinc-blende and the diamond lattices. In what follows we adopt the cubic approximation given by Eq.~(\ref{nuli}).

\subsection{Cylindrical coordinates}\label{III_cyl}

For the Bloch function $u_c$ is invariant under rotation of the Cartesian coordinate system around the $z$-axis, the procedure of transformation of the Kane Hamiltonian to cylindrical coordinates trivially follows the one for the valence-band case. We accordingly introduce the rotationally invariant functions $ \widetilde {\mathbf u}' = \left(u_c\ u_r\ u_\phi\ u_z \right)$, so that $\mathbf u' =  \widetilde{\mathbf u}' \mathbf S'$, where the unitary matrix
\begin{equation}
\label{Sprime}
\mathbf S' =
\begin{pmatrix}
1 & 0& 0& 0\\
0& \cos \phi & \sin \phi & 0\cr
0&-\sin \phi & \cos \phi & 0\cr
0&0& 0 & 1
\end{pmatrix}
\end{equation}
plays the role of the matrix $\mathbf S$ for the valence-band case, see Eq.~(\ref{S}). To avoid the $\phi$-dependence in the spin-orbit Hamiltonian after the transformation realized with the matrix $\mathbf S'$, we use the unitary transformation defined by the matrix $\widetilde {\mathbf S}'$,
\begin{equation}\label{tildeSprimeexplicit}
\widetilde {\mathbf S}' =
\begin{pmatrix}
\mathrm e ^{ i \frac \phi 2 \sigma_z }&0&0&0\\
0& \mathrm e ^{ i \frac \phi 2 \sigma_z }&0&0\cr
0&0&\mathrm e ^{ i \frac \phi 2 \sigma_z }&0\cr
0&0&0&\mathrm e ^{ i \frac \phi 2 \sigma_z }
\end{pmatrix}.
\end{equation}
The resulting doubly transformed Hamiltonian $\mathbf H '$ of Eq.~(\ref{Schkane}) is
\begin{equation}\label{Hprimetransformed}
\widetilde {\widetilde {\mathbf H}}{}' = \widetilde {\mathbf S}' {\mathbf S}'
{\mathbf H}' {\mathbf S'}{}^{-1} \widetilde {\mathbf S}'{}^{-1} =
\begin{pmatrix}
\widetilde {\widetilde H}_c & \widetilde {\widetilde {\mathbf H}}_{cv}\cr
\widetilde {\widetilde {\mathbf H}}^\dag _{cv}& \widetilde {\widetilde {\mathbf H}}_v
\end{pmatrix},
\end{equation}
where
\begin{equation}
\begin{split}
\widetilde {\widetilde H}_c = & - A'_1 \nabla_z^2 - A'_2 \left[ \nabla_r^2 + \frac 1 r \nabla_r + \frac 1 {r^2} \widehat {\boldsymbol \nabla}_\phi^2  \right] \\
&+U_c +a_1 \varepsilon_{zz} + a_2  \left( \varepsilon_{rr} + \varepsilon_{\phi\phi}\right),
\end{split}
\end{equation}
\begin{equation}\label{Hcvtilde}
\widetilde {\widetilde {\mathbf H}}_{cv} = \begin{pmatrix}
\widetilde {\widetilde H}_{cv1} & \widetilde {\widetilde H}_{cv2} & \widetilde {\widetilde H}_{cv3} \end{pmatrix},
\end{equation}
with
\begin{equation}\label{Hcxtilde}
\widetilde {\widetilde H}_{cv1} = P_2 \left( \nabla_r + \frac 1 r \right),
\end{equation}
\begin{equation}\label{Hcyztilde}
\widetilde {\widetilde H}_{cv2} = P_2 \frac 1  r \widehat {\boldsymbol \nabla}_\phi,\quad
\widetilde {\widetilde  H}_{cv3} = P_1 \nabla_z.
\end{equation}
Note that
\begin{equation}\label{Hcxtildedag}
\widetilde {\widetilde H}{}^\dag_{cv1} = - P_2 \nabla_r,
\end{equation}
while
\begin{equation}
\widetilde {\widetilde H}{}^\dag_{cv2} = - P_2 \frac 1  r \widehat {\boldsymbol \nabla}_\phi , \quad
\widetilde {\widetilde  H}{}^\dag_{cv3} = -P_1 \nabla_z.
\end{equation}
For the valence band block $\widetilde {\widetilde {\mathbf H}}_v$ we have
\begin{equation}\label{Hvtilde}
\widetilde {\widetilde {\mathbf H}}_v = \mathbf H^{(0)} + \mathbf H^{(\sigma)} + \mathbf H^{(\varepsilon)} +\widetilde {\widetilde {\mathbf H}}{}^{\prime (k)}, 
\end{equation}
where $\mathbf H^{(0)}$, $\mathbf H^{(\sigma)}$ and $\mathbf H^{(\varepsilon)}$ are given by Eqs.~(\ref{H0}), (\ref{Hso}) and (\ref{tildeHe}), respectively. For the kinetic energy part $\widetilde {\widetilde {\mathbf H}}{}^{\prime (k)}$ we use the matrix Hamiltonian of Eq.~(\ref{tildeHk}) with the substitute $\nabla _\phi \rightarrow \widehat {\boldsymbol \nabla}_\phi$, see Eq.~(\ref{nabla-bold}), along with the substitutes $L_1 \rightarrow  L'_1$, $L_2 \rightarrow  L'_2$ $N_1 \rightarrow  N'_1$ and $N_2 \rightarrow  N'_2$, see Eqs.~(\ref{LNprimes1}), (\ref{LNprimes2}), and (\ref{LNprimes3}).

For cylindrically symmetric structures, the Hamiltonian $\widetilde {\widetilde {\mathbf H}}{}'$ commutes with the operator $-i\nabla_\phi$. The EFs $\widetilde {\widetilde {\mathbf F}}{}' = \widetilde {\mathbf S}' \mathbf S' \mathbf F '$, which satisfy the system of equations
\begin{equation}\label{Kane_cyl}
\widetilde {\widetilde {\mathbf H}}{}' \widetilde {\widetilde {\mathbf F}}{}' = E \widetilde {\widetilde {\mathbf F}}{}',
\end{equation}
can be chosen as follows:
\begin{equation} \label{tildetildeFKane}
\widetilde {\widetilde {\mathbf F}}{}' = \frac {{\mathrm e}^{im\phi}}{\sqrt {2\pi}}\, \mathbf f' \left(r, z \right),
\end{equation}
with $m = \pm 1/2, \pm 3/2, \dots$, while the functions $\mathbf f' \left(r, z \right) = \left( \mathbf f_c \ \mathbf f_1 \ \mathbf f_2 \ \mathbf f_3 \right)^{\mathrm T}$, where 
$\mathbf f_j = \left( f^{(u)}_j \ f^{(d)}_j \right)^{\mathrm T}$, $j=c,1,2,3$,
do not depend on $\phi$.
\begin{table}[tb]
\squeezetable
\caption{Boundary conditions for the conduction-band component of $\mathbf f'$ at $r =0$ for different $m$.}
\begin{ruledtabular}
\begin{tabular}{lccc}
$m$ & 1/2 & --1/2 & other half-integers\\
\hline
$f^{(u)}_c$           &  --- &  0      & 0\\
$\nabla _r f^{(u)}_c$ &  0   &  ---    & ---\\
$f^{(d)}_c$           &  0   &  ---    & 0\\
$\nabla _r f^{(d)}_c$ &  --- &  0      & ---
\end{tabular}
\end{ruledtabular}
\label{BC_cond}
\end{table}

The BCs for the valence-band components of $\mathbf f'$ on the symmetry axis $r=0$ are the same as those given in Table~\ref{BC6x6}. The relations of the conduction-band elements of $\mathbf f'$ to the old EFs $\mathbf F '$,
\begin{equation}
F^{(u)}_c = \frac {\mathrm e^{i\left( m - 1/2\right) \phi} } {\sqrt {2\pi}}\,f^{(u)}_c, \quad
F^{(d)}_c = \frac {\mathrm e^{i\left( m + 1/2\right) \phi} } {\sqrt {2\pi}}\,f^{(d)}_c,
\end{equation}
formally coincide with such for the valence-band elements $f^{(u)}_3$ and $f^{(d)}_3$, see Eq.~(\ref{u_components}), resulting in the same BCs at $r=0$, which are presented in Table~\ref{BC_cond}. The superfluity of some of the BCs, shown in Table~\ref{BC6x6} in parentheses, is easily demonstrated, as in the valence-band case, from the analysis of the second and third equations in Eq.~(\ref{Kane_cyl}), using also the corresponding BCs for the conduction-band components of $\mathbf f'$.

The external magnetic field $B_z$ is treated analogously to the valence-band case, see Appendix~\ref{II_mf}, through the substitutes given by Eqs.~(\ref{nablas}) and (\ref{cyl-momentum}), including also the following Pauli term $\mathbf H ^{\prime (B)}$ in the Hamiltonian $\widetilde {\widetilde {\mathbf H}}{}'$:
\begin{widetext}
\begin{equation}\label{HBKane}
\mathbf H ^{\prime (B)} = \begin{pmatrix}
\frac {g_0 \mu_B}2 \sigma_z B_z & 0 &0&0\cr
0&\frac {g_0 \mu_B}2 \sigma_z B_z & -\frac {ie}{2\hbar c}Q B_z &0\cr
0&\frac {ie}{2\hbar c}Q B_z & \frac {g_0 \mu_B}2 \sigma_z B_z & 0\cr
0&0&0&\frac {g_0 \mu_B}2 \sigma_z B_z
\end{pmatrix},
\end{equation}
\end{widetext}
which is invariant under the transformations realized with the matrices $\mathbf S'$ and $\widetilde {\mathbf S}'$.

For zinc-blende materials, we use the relations given in Sec.~\ref{II_zb}, along with the following ones: $a_1 = a_2$, $m_1 = m_2$, $P_1 = P_2$, and $E_{v1} = E_{v6}$. 

\section{Valence-band states in wurtzite QWs}\label{IV_holes}

Actual heterostructures, such as wurtzite GaN/AlN QDs of truncated pyramidal geometry with a hexagonal base \cite{Arley} or zinc-blende InAs/GaAs square-base pyramidal QDs, \cite{Bruls} are devoid of the cylindrical symmetry. To take advantages of the cylindrical coordinate representation of the EM equations, one must first assume the cylindrical symmetry approximation for the electron systems of interest. Zinc-blende materials should be of special concern here due to the necessity to accept the cylindrical geometry, spherical band and spherical deformation potentials approximations simultaneously. On the other hand, one may expect that wurtzite structures grown along the [0001] direction are very close to cylindrically symmetric because of the cylindrically symmetric band structure, in the EM approximation including the Kane model, and the argument that a regular hexagon is geometrically close to a circle.

\begin{figure}[tb]
\includegraphics [width=5cm]{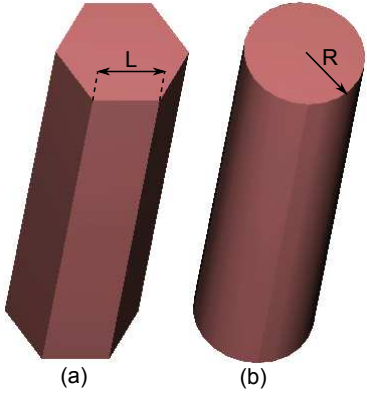}
\caption{(Color online) Two studied types of QWs with the same cross-sectional area: (a) hexagonal QW with the hexagon side of length $L$, (b) cylindrical QW with the radius $R = L \sqrt{3\sqrt 3/2\pi}$.}
\label{2wires}
\end{figure}

To verify this assumption, we compute the spectrum of the valence-band states, as a function of the wavenumber $k_z$, in an isolated [0001] wurtzite GaN QW that has a regular hexagonal cross-section. We use the valence-band EM equations in Cartesian coordinates, see Eq.~(\ref{H}). We also adopt the cylindrically symmetric geometry approximation for the QW, preserving the same cross-sectional area, see Fig.~\ref{2wires}, and compute the spectrum by using Eq.~(\ref{Hf_Ef}) and the BCs from Table~\ref{BC6x6}. For both cases the EFs are set to zero at the GaN-vacuum boundary.
\begin{figure}[tb]
\includegraphics [width=8.5cm]{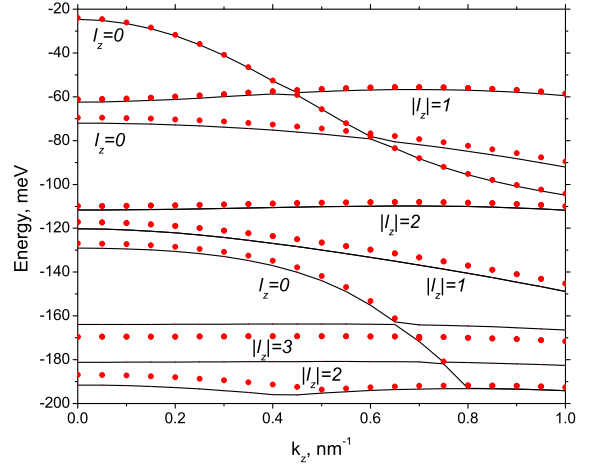}
\caption{(Color online) Valence-band states in a hexagonal GaN QW with the length of the hexagon side $L =2.2$~nm, solid lines; valence-band states in a cylindrical GaN QW with the radius $R =2$~nm, characterized by the projection $l_z$ of orbital momentum on the $z$-axis, dotted lines.}
\label{spectra}
\end{figure}
\begin{figure}[tb]
\includegraphics [width=8.5cm]{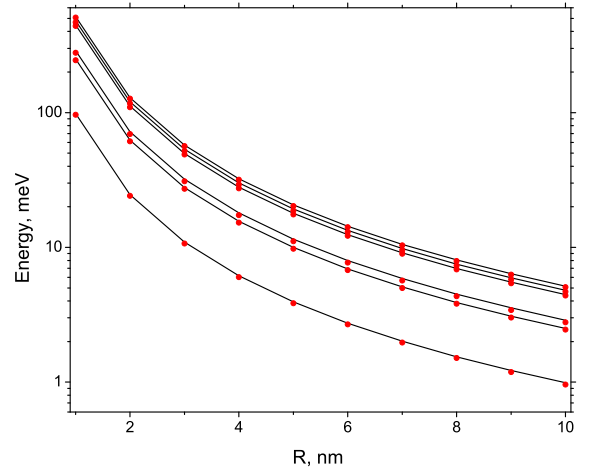}
\caption{(Color online) Absolute values of the energy of several lowest valence-band states at $k_z =0$ in a hexagonal GaN QW, solid lines, and in a cylindrical GaN QW, dots, as a function of the QW radius R; the corresponding hexagon's side length is $L=1.1R$.}
\label{radius-depend}
\end{figure}
The following material parameters are used. The radius of the QW in the cylindrical approximation is $R =2$~nm. The same cross-sectional area has a hexagonal QW with the hexagon side of length $L = R \sqrt{2\pi/3\sqrt 3} \approx 1.100 R = 2.2$~nm. The band structure parameters of wurtzite GaN are taken from Ref.~\onlinecite{Vurgaftman_old}: $A_1 = -7.21$, $A_2=-0.44$, $A_3=6.68$, $A_4=-3.46$ $A_5=-3.40$, $A_6=-4.90$. The strain Hamiltonian gives no contribution in isolated QWs. For simplicity, we neglect the spin-orbit interaction, $\Delta_2 = \Delta_3 =0$, and also the crystal splitting $\Delta_1 =0$ for consistency, so that $E_{v1} = E_{v6}$, with the value being used as the zero-energy reference point, and put $A_7 =0$. The upper and the down spinor elements of the functions $\mathbf f$, see Eq.~(\ref{Hf_Ef}), are then not coupled, and the states are characterized by the projection $l_z$ of orbital momentum on the $z$-axis, with $l_z = m-1/2$ for the upper spinor elements of $\mathbf f$, and $l_z = m+1/2$ for the down ones.

Due to properties of the space group $C_{6v}$, all states in the hexagonal QW are either non-degenerate or have a two-fold degeneracy, not considering the spin degeneracy. One might expect that the non-degenerate states correspond to the states with $l_z =0$ for the cylindrical QW, while the two-fold degenerate states are related to the states $l_z =\pm 1,\pm 2, \ldots$ in the cylindrical QW. As can be seen in Table~\ref{d} and Fig.~\ref{spectra}, the ground and several excited states in these QWs not only have similar behavior but are also very close in energy, with the difference being less than $4\%$. An exception holds for the states with $\vert l_z \vert = 3n$, where $n=1,2,\ldots $, which turn into non-degenerate states in the hexagonal QW, see Appendix~\ref{symmetry_groups}. The smallness of the corresponding energy splittings may be established as a criterion for applicability of the cylindrical symmetry approximation for hexagonal systems. It is satisfied for the lowest states with $\vert l_z \vert = 3$, see Table~\ref{d} and Fig.~\ref{spectra}, for the hexagonal splitting is only $10\%$ of the energy of the states.

The results remain even quantitatively the same for other actual \cite{Persson} sizes of the QWs. The energy of the valence-band states in hexagonal and equivalent cylindrical QWs is shown in Fig.~\ref{radius-depend} for $k_z =0$ as a function of the radius $R$, from $R=1$~nm to $R=10$~nm with the step $\delta R = 1$~nm (the values at $R=2$~nm corresponds to the detailed data in Table~\ref{d} and Fig.~\ref{spectra}). Shown are only several lowest states that are equivalent to the ones with $\vert l_z \vert \le 2$ for cylindrical geometry. The fit of the spectrum in cylindrical QWs to that in hexagonal QWs is excellent, the energy difference ranging from $1.6\%$ to $3.7\%$. The hexagonal symmetry splitting of the lowest degenerate state with $\vert l_z \vert = 3$ (not shown) is still about $10\%$ of the energy of the state, experiencing a variation with $R$ only in the third significant digit.

\begin{table}[tb]
\squeezetable
\caption{Spectrum $E^{(h)}_n$ of several valence-band states (each doubly spin-degenerate) in a wurtzite GaN QW with the regular hexagonal (side length 2.2~nm) cross-section and corresponding energies $E^{(c)}$ of the states in the QW with the circular (radius 2~nm) cross-section for two values of the wavenumber $k_z$, presented in Fig.~\ref{spectra}. The spectrum $E^{(c)}$ is characterized by the projection $l_z$ of angular momentum on the symmetry axis.}
\begin{ruledtabular}
\begin{tabular}{lccccc}
$k_z$, nm$^{-1}$ & 0 & 1\\
\hline
$E^{(h)}_1$    ($E^{(c)}$), meV    &  --24.63 (--24.09\footnotemark[1])  &  --59.48 (--58.49\footnotemark[2]) \\
$E^{(h)}_2$    ($E^{(c)}$), meV    &  --62.41 (--61.15\footnotemark[2])  & --59.48 (--58.49\footnotemark[2])\\
$E^{(h)}_3$    ($E^{(c)}$), meV    &  --62.41 (--61.15\footnotemark[2])  & --92.06 (--89.55\footnotemark[1])\\
$E^{(h)}_4$    ($E^{(c)}$), meV    &  --72.03 (--69.48\footnotemark[1])  & --104.90 (--104.17\footnotemark[1])\\
$E^{(h)}_5$    ($E^{(c)}$), meV    &  --111.70 (--109.84\footnotemark[3]) & --111.72 (--109.93\footnotemark[3])\\
$E^{(h)}_6$    ($E^{(c)}$), meV    &  --111.70 (--109.84\footnotemark[3]) & --111.72 (--109.93\footnotemark[3])\\
$E^{(h)}_7$    ($E^{(c)}$), meV    &  --120.23 (--117.16\footnotemark[2]) & --148.82 (--145.22\footnotemark[2])\\
$E^{(h)}_8$    ($E^{(c)}$), meV    &  --120.23 (--117.16\footnotemark[2]) & --148.82 (--145.22\footnotemark[2])\\
$E^{(h)}_9$    ($E^{(c)}$), meV    &  --129.04 (--126.9\footnotemark[1]) & --166.49 (--171.55\footnotemark[4])\\
$E^{(h)}_{10}$ ($E^{(c)}$), meV    &  --163.93 (--169.53\footnotemark[4]) & --182.65 (--171.55\footnotemark[4])\\
$E^{(h)}_{11}$ ($E^{(c)}$), meV    &  --181.17 (--169.53\footnotemark[4]) & --194.11 (--192.58\footnotemark[3])\\
$E^{(h)}_{12}$ ($E^{(c)}$), meV    &  --191.58\footnotemark[5] (--186.89\footnotemark[3]) & --194.11 (--192.58\footnotemark[3])
\end{tabular}
\end{ruledtabular}
\footnotetext[1]{$l_z = 0$.}
\footnotetext[2]{$\vert l_z  \vert= 1$.}
\footnotetext[3]{$\vert l_z\vert = 2$.}
\footnotetext[4]{$\vert l_z\vert = 3$.}
\footnotetext[5]{Degenerate with the $E^{(h)}_{13}$ level.}
\label{d}
\end{table}

\section{Discussion}\label{V_discuss}

The approach of Sercel and Vahala, \cite{SV} proposed more than two decades ago, allows to reduce the number of independent variables in the multiband Hamiltonians for radially symmetric structures by integrating out angular degrees of freedom. Being a development of the ideas of Baldereschi and Lipari,\cite{Baldereschi} it has still inherited an indefinite status of the BCs for the EFs at the origin, for spherical coordinates, or on the symmetry axis, for cylindrical ones. In particular, Sercel and Vahala did not derive the BCs but suggested instead that the EFs be regular. By definition, a regular at a point $\xi=\xi_0$ function $f(\xi)$ is presented by a convergent power series
\begin{equation}\label{regular}
f(\xi)=\sum_{n=0}^{\infty}\left(\xi-\xi_0\right)^n f_n,
\end{equation}
with some operators $f_n$, for $\vert \xi-\xi_0\vert < \epsilon$ in some vicinity $\epsilon >0$ of the point $\xi=\xi_0$. \cite{Korn} However, in cylindrical or spherical coordinates, the EFs as functions of a radial variable $\xi$ are formally not regular at $\xi =0$ because they are not defined for $\xi <0$. As a consequence, the BCs are not specified by the regularity arguments; the formalism of Sercel and Vahala is incomplete and cannot be put into practice. Neither this fact has been conceived nor an alternative radially symmetric coordinate representation for multiband Hamiltonians has been proposed in the literature up to now. We note here that the BCs for the EFs in cylindrical or spherical coordinates may not be conventional, as seen in Table~\ref{BC6x6}, and may not correspond to a hypothetical situation when each EF component has either even or odd parity. \cite{incorrectBC}

We have not found the BCs to complement the formalism of Sercel and Vahala, but developed a different one in view of the following. For a cylindrically symmetric nanostructure, for example, the effective Hamiltonian for each eigenvalue of the operator of projection of total angular momentum on the symmetry axis is to be derived individually in the approach of Sercel and Vahala. Reminding the formulation of Heisenberg's matrix mechanics, \cite{Messiah} such a procedure is quite inefficient. We should admit that this may not be a serious shortcoming because in most practical cases properties of only several low-lying states are of interest, while states with large values of angular momentum are usually characterized by large values of eigenenergy, which may even fall beyond the range of applicability of the EM equations. Nevertheless, we had to analyze states with $\vert l_z \vert = 3$ to grasp the effect of their splitting in actual QWs with hexagonal symmetry. Hamiltonian for such values of the orbital momentum, albeit for zinc-blende systems, has not been presented previously in an explicit form.

The cylindrical coordinate representation is directly applicable only to idealized structures such as zinc-blende (truncated) conical QDs and cylindrical QWs, in the spherical band approximation, and cylindrically symmetric wurtzite structures grown along the [0001] crystallographic direction. An unbounded cylindrically symmetric 2D or 3D electron system can also be such an object if a particular task requires quantum numbers of the angular momentum operator rather than momentum. A particular study is yet to confirm the applicability of the cylindrical symmetry approximation for the whole range of heterostructures possessing different geometry, size, and material composition. We have presented the analysis only for the simplest but actual for applications \cite{Persson} case of the valence-band states in isolated [0001] wurtzite GaN QWs with a regular hexagonal cross-section. We have obtained evidences that such systems can be approximated by cylindrical QWs. The excellent fit of the spectrum of states in cylindrical QWs to that in hexagonal QWs also indicates that one may expect qualitatively similar picture when actual \cite{Arley} wurtzite QDs are approximated by cylindrically symmetric ones.

Analogously, the multiband Hamiltonians can also be expressed in spherical coordinates. While spherical QDs are occasionally considered in the literature, \cite{globe1,globe2} such geometry is presumably too crude an approximation for actual truncated pyramidal QDs, and the spherical coordinate representation would be of very limited interest.

\section{Conclusions}\label{V_concl}

The goal of this work was to find an efficient scheme to express the valence-band and Kane envelope-function Hamiltonians for wurtzite and zinc-blende semiconductor heterostructures in cylindrical coordinates.  Such a representation can considerably reduce the computational cost and facilitate analyzing the electron states in ultimate low-dimensional electron systems---QWs and especially QDs---if the systems can be approximated as cylindrically symmetric. To achieve the goal, we have constructed rotationally invariant basis functions that allow us to use the corresponding EFs as eigenstates of the operator of projection of total angular momentum on the symmetry axis. As a result, our multiband Hamiltonians depend on a single parameter, which is an eigenvalue of the operator of projection of total angular momentum on the symmetry axis. Such representation is conventional, reminding the textbook case of the usual Schr\"odinger equation for a cylindrically symmetric system. \cite{Messiah,Landavshitz-QM} In the Hamiltonians, we have taken into account the deformation effects and have made an allowance for an external magnetic field applied along the symmetry axis. We have supplemented the multiband EM equations by BCs imposed on the EFs on the symmetry axis, thus making the cylindrical coordinate formalism complete.

The obtained results have been applied to analyze the valence-band states in isolated wurtzite GaN QWs, grown along the [0001] crystallographic direction, with a regular hexagonal cross-section. We have revealed a good correspondence between the spectrum in such QWs and a spectrum of the states in equivalent cylindrical QWs. The degenerate states with $\vert l_z \vert = 3,6,9,\ldots$ in cylindrical QWs are equivalent to split states in hexagonal QWs. We have suggested smallness of the splitting energy as a criterion for applicability of the cylindrical symmetry approximation for hexagonal systems.

\section*{Acknowledgments}

The author is grateful to COMSOL, Inc. for providing a copy of COMSOL Multiphysics 4.2a at COMSOL Workshop, University of Waterloo, Ontario, Canada. A part of this work has been done at M$^2$NET Laboratory, Wilfrid Laurier University, Ontario, Canada.

\appendix

\section{External magnetic field}\label{II_mf}

To take into account an external magnetic field $B_z$, we use the symmetric gauge for the vector potential $\mathbf A = \left( A_x,A_y,A_z \right)= \left( -yB_z/2,xB_z/2,0\right)$ in Cartesian coordinates, or $\left( A_r,A_\phi,A_z \right) = \left(0,r B_z/2, 0\right)$ in cylindrical coordinates. We make the following substitutes in the Hamiltonian of Eq.~(\ref{Hk}):
\begin{equation}
k_j  \rightarrow K_j \equiv k_j + \frac e {\hbar c} A_j, \quad j = x,y,z,
\end{equation}
where $-e<0$ is the electron charge, and $c$ is the speed of light in vacuum. All non-commutative products should be replaced by the symmetrized products, \cite{Luttinger2}
\begin{equation}
K_x K_y \rightarrow \frac 1 2 \{ K_x,K_y \}_+ \equiv \frac 1 2 \left( K_xK_y + K_yK_x \right).
\end{equation}
In addition, the Hamiltonian $\mathbf H$ of Eq.~(\ref{H}) must include the effective Pauli term $\mathbf H^{(B)}$,
\begin{equation}\label{HB}
\mathbf H ^{(B)} = \begin{pmatrix}
\frac {g_0 \mu_B}2 \sigma_z B_z & -\frac {ie}{2\hbar c}Q B_z &0\cr
\frac {ie}{2\hbar c}Q B_z & \frac {g_0 \mu_B}2 \sigma_z B_z & 0\cr
0&0&\frac {g_0 \mu_B}2 \sigma_z B_z
\end{pmatrix},
\end{equation}
where $g_0 \approx 2 $ is the free electron $g$-factor, $\mu_B$ is the Bohr magnetron, and $Q$ is a material parameter, \cite{Luttinger2}
\begin{equation}\label{Q}
Q= \frac {\hbar^2}{m_0^2}\sum_\gamma \frac {p^{(x)}_{x\gamma}p^{(y)}_{\gamma y} - p^{(y)}_{x\gamma}p^{(x)}_{\gamma y}} {E_{v6} - E_\gamma},
\end{equation}
where $E_{v6}$ and $E_\gamma$ are the energies of the valence $\Gamma_6$ state and the state with the index $\gamma$, respectively, of the reference material; $p^{(j)}_{j '\gamma}$ are the matrix elements of the momentum operator between the zone-center Bloch functions $u_\gamma$ and $u_{j '}$, where $j,j '=x,y$, that is $p^{(j)}_{j ' \gamma} = \langle u_{j '} \vert \hbar k_j \vert u_\gamma \rangle$.

Let us compare Eq.~(\ref{Q}) with a similar expression \cite{Chuang} for the material parameter $N_1$,
\begin{equation}\label{N1}
N_1= \frac {\hbar^2}{m_0^2}\sum_\gamma \frac {p^{(x)}_{x\gamma}p^{(y)}_{\gamma y} + p^{(y)}_{x\gamma}p^{(x)}_{\gamma y}} {E_{v6} - E_\gamma},
\end{equation}
and note that $p^{(y)}_{x\gamma} = 0$ for the nearest $\gamma$ bands, which belong to the representation $\Gamma_1$. This indicates that the approximation $Q \approx N_1$ may be satisfactory.

In cylindrical coordinates, we symmetrize the products of the operators $\nabla_\phi$ and $\nabla_r$ entering the Hamiltonian of Eq.~(\ref{tildeHk}),
\begin{equation}\label{nablas}
\nabla_\phi \nabla_r \rightarrow  \frac 1 2 \{ \nabla_\phi,\nabla_r \}_+,
\end{equation} 
and make the following substitute:
\begin{equation}\label{cyl-momentum}
\nabla_\phi \rightarrow \nabla_\phi + \frac {ie} {2\hbar c} B_z r^2,
\end{equation} 
so that, for example,
\begin{widetext}
\begin{equation}\label{H12-B}
\widetilde {H}^{(k)}_{12} =
- \left\{ \nabla_\phi + \frac {ie} {2\hbar c} B_z r^2, \nabla_r \right\}_+ \frac {L_1} {2r} + \frac {M_1} {2r} \left\{ \nabla_\phi + \frac {ie} {2\hbar c} B_z r^2, \nabla_r \right\}_+ + \frac {M_1} {r^2} \left( \nabla_\phi + \frac {ie} {2\hbar c} B_z r^2\right).
\end{equation}
\end{widetext}
The rest arguments directly follow the ones for the case without magnetic field. After the unitary transformations, realized with the operators $\mathbf S$ and $\widetilde {\mathbf S}$, the effective Pauli term of Eq.~(\ref{HB}) remains invariant, 
\begin{equation}\label{Pauli-transformed}
\widetilde {\widetilde {\mathbf H}}\,\!^{(B)} = \widetilde {\mathbf S} {\mathbf S}
{\mathbf H}^{(B)} {\mathbf S}^{-1} \widetilde {\mathbf S} ^{-1} = {\mathbf H}^{(B)}.
\end{equation}
The resulting full Hamiltonian commutes with the operator $-i\nabla_\phi$, if the system is cylindrically symmetric. The EFs have the form of Eq.~(\ref{tildetildeF}), where the functions $\mathbf f$ are obtained by solving the following system of equation:
\begin{equation}\label{shroedinger_mf}
\left( \mathbf H^{(0)} + {\mathbf H}^{(\sigma)} + \widetilde {\mathbf H}^{(\varepsilon)} + {\mathbf H}^{(B)} +\widetilde {\widetilde {\mathbf H}}\,\!^{(k)}_m \right) \mathbf f = E \mathbf f,
\end{equation}
with the kinetic energy Hamiltonian $\widetilde {\widetilde {\mathbf H}}\,\!^{(k)}_m$ being presented by the matrix operator
\begin{widetext}
\begin{equation} \label{tildeHkmB}
\begin{split}
\begin{bmatrix}
M_1 \frac {\hat {\mathbf m}^2} {r^2} -
L_1 \left[ \nabla_r^2 + \nabla_r \frac 1 r \right] 
- M_2 \nabla_z^2 &
i M_1 \left[ \frac  1 {2r} \{ {\hat {\mathbf m}}, \nabla_r \}_+ +\frac {\hat {\mathbf m}} {r^2} \right]
-i \{ {\hat {\mathbf m}}, \nabla_r\}_+ \frac {L_1} {2r} &
- N_2 \nabla_r \nabla_z + i N_3 \nabla_r \\
i \{ {\hat {\mathbf m}}, \nabla_r\}_+ \frac {M_1} {2r} 
- i L_1 \left[ \frac  1 {2r} \{ {\hat {\mathbf m}}, \nabla_r\}_+ +\frac {\hat {\mathbf m}} {r^2} \right]&
L_1 \frac {\hat {\mathbf m}^2} {r^2} -
M_1 \left[ \nabla_r^2 + \nabla_r \frac 1 r \right] - M_2 \nabla_z^2 &
-i N_2 \frac {\hat {\mathbf m}} r \nabla_z - N_3 \frac {\hat {\mathbf m}} r\\
- N_2 \left[ \nabla_r +\frac 1 r \right]\nabla_z -i N_3 \left[ \nabla_r +\frac 1 r \right]&
- i N_2 \frac {\hat {\mathbf m}} r \nabla_z + N_3 \frac {\hat {\mathbf m}} r&
M_3 \left[ \frac {\hat {\mathbf m}^2} {r^2} - \nabla_r^2 - \frac 1 r \nabla_r \right]-
L_2 \nabla_z^2
\end{bmatrix},
\end{split}
\end{equation}
\end{widetext}
where the matrix operator $\hat {\mathbf m}$ is 
\begin{equation}
\hat {\mathbf m} = 
\begin{pmatrix}
\frac {e} {2\hbar c} B_z r^2 + m - \frac 1 2 &0\cr
0& \frac {e} {2\hbar c} B_z r^2 + m +\frac 1 2
\end{pmatrix},
\end{equation}
and $\mathbf H^{(0)}$, ${\mathbf H}^{(\sigma)}$, $\widetilde {\mathbf H}^{(\varepsilon)}$, and $\mathbf H^{(B)}$ are given by Eqs.~(\ref{H0}), (\ref{Hso}), (\ref{tildeHe}), and (\ref{HB}), respectively, while taking notice of Eq.~(\ref{strain_cyl_sym}). Also holds Eq.~(\ref{psi2}). 

\section{Symmetry groups $C_{6v}$ and $C_{\infty v}$}\label{symmetry_groups}

The states in cylindrical and hexagonal QWs are classified by the irreducible representations of the axial symmetry group $C_{\infty v}$ and the group $C_{6v}$, respectively, \cite{Landavshitz-QM} characters of which are given in Tables~\ref{Cinfv} and \ref{C6v}.
\begin{table}[tbh]
\squeezetable
\caption{Characters of the irreducible representations of the group $C_{\infty v}$.}
\begin{ruledtabular}
\begin{tabular}{lrrr}
 & $E$ & $2C(\phi)$ & $\infty \sigma_v$ \\
\hline
$A_1$ &1&1&1 \\
$A_2$ &1&1&--1\\
$E_l$ &2&$2\cos{l\phi}$&0
\end{tabular}
\end{ruledtabular}
\label{Cinfv}
\end{table}
\begin{table}[tbh]
\squeezetable
\caption{Characters of the irreducible representations of the group $C_{6v}$.}
\begin{ruledtabular}
\begin{tabular}{lrrrrrr}
 & $E$ & $C_2$ & $2C_3$ & $2C_6$ & $3\sigma_v$ & $3\sigma^\prime_v$ \\
\hline
$A'_1$ &1&1&1&1&1&1 \\
$A'_2$ &1&1&1&1&--1&--1 \\
$B'_1$ &1&--1&1&--1&--1&1 \\
$B'_2$ &1&--1&1&--1&1&--1 \\
$E'_1$ &2&--2&--1&1&0&0 \\
$E'_2$ &2&2&--1&--1&0&0
\end{tabular}
\end{ruledtabular}
\label{C6v}
\end{table}
\begin{table}[tbh]
\squeezetable
\caption{Characters of the representations of the group $C_{\infty v}$ for the operations of the symmetry group $C_{6v}$.}
\begin{ruledtabular}
\begin{tabular}{lrrrrrr}
 & $E$ & $C_2$ & $2C_3$ & $2C_6$ & $3\sigma_v$ & $3\sigma^\prime_v$ \\
\hline
$A_1$ &1&1&1&1&1&1 \\
$A_2$ &1&1&1&1&--1&--1 \\
$E_1,\ E_5,\ E_7,\ E_{11},\ E_{13}, \ldots $ &2&--2&--1&1&0&0 \\
$E_2,\ E_4,\ E_8,\ E_{10},\ E_{14}, \ldots $ &2&2&--1&--1&0&0 \\
$E_3,\ E_9,\ E_{15},\ E_{21},\ E_{27}, \ldots $ &2&--2&2&--2&0&0 \\
$E_6,\ E_{12},\ E_{18},\ E_{24},\ E_{30}, \ldots $ &2&2&2&2&0&0
\end{tabular}
\end{ruledtabular}
\label{Cinfv_6v}
\end{table}
In Table~\ref{Cinfv_6v}, we also present characters of the representations of the group $C_{\infty v}$ for the operations of the symmetry group $C_{6v}$ to obtain the following decomposition rules:
\begin{equation}
\begin{split}
&A_1 = A'_1, \quad A_2 = A'_2, \quad E_{l_1} = E'_1, \quad E_{l_2} = E'_2,\\
&E_{l_3} = B'_1 + B'_2, \quad E_{l_4} = A'_1+A'_2, 
\end{split}
\end{equation}
where
$l_1$ is an odd positive integer not multiple of $3$, $l_2$ is an even positive integer not multiple of $3$, $l_3$ is a positive integer multiple of $3$ but not multiple of $6$, and $l_4$ is a positive integer multiple of $6$. The decomposition rules indicate that the hexagonal symmetry lifts the orbital degeneracy only for the states with projection of angular momentum on the symmetry axis $\vert l_z \vert = 3n$, where $n$ is a positive integer. This symmetry analysis is also valid for wurtzite quantum dots with a regular hexagonal cross-section.


\begin{thebibliography}{99}

\bibitem{QD} E.~Borovitskaya and M.E.~Shur, in {\em Quantum Dots}, Selected Topics in Electronics and Systems Vol.~{\bf 25}, edited by E.~Borovitskaya and M.E.~Shur (World Scientific, Singapore, 2002), p.~1.

\bibitem{Kapon} E.~Kapon, in {\em Quantum Well Lasers}, edited by: P.S.~Zory, Jr. (Academic Press, New York, 1993), p. 461.

\bibitem{Petroff} P.M.~Petroff, A.~Lorke, and A.~Imamoglu, Physics Today {\bf 54}(5), 46 (2001).

\bibitem{Gudiksen} M.S.~Gudiksen, L.J.~Lauhon, J.~Wang, D.C.~Smith, and C.M.~Lie\-ber, Nature (London) {\bf 415}, 617 (2002).

\bibitem{Anikeeva} P.O.~Anikeeva, J.E.~Halpert, M.G.~Bawendi, and V.~Bulovi\'{c}, Nano Letters {\bf 9} 2532 (2009).

\bibitem{Hawrylak_QC} J.A.~Brum and P.~Hawrylak, Superlattices Microstruct. {\bf 22}, 431 (1997).

\bibitem{Loss} D.~Loss and D.P.~DiVincenzo, Phys.\ Rev.\ A {\bf 57}, 120 (1998).

\bibitem{Elzerman} J.M.~Elzerman, R.~Hanson, L.H.W.~van~Beveren, S.~Tarucha,
L.M.K.~Vandersypen, and L.P.~Kouwenhoven, in {\em Quantum Dots: A Doorway to Nanoscale Physics}, Lecture Notes in Physics Vol.~{\bf 667}, edited by W.D.~Heiss (Springer, Berlin, 2005), p. 25.

\bibitem{QC_holes} B.D.~Gerardot, D.~Brunner, P.A.~Dalgarno, P.~\"{O}hberg, S.~Seidl, M.~Kroner, K.~Karrai, N.G.~Stoltz, P.M.~Petroff, and R.J.~Warburton, Nature (London) {\bf 451}, 441 (2008).

\bibitem{Zunger_qd} H.~Fu, L.-W.~Wang, and A.~Zunger, Phys.\ Rev.\ B {\bf 57}, 9971 (1998).

\bibitem{DiCarlo} A.~Di~Carlo, Semicond.\ Sci.\ Technol.\ {\bf 18} R1 (2003).

\bibitem{Persson} M.P.~Persson and A.~Di~Carlo, J.\ Appl.\ Phys. {\bf 104}, 073718 (2008).

\bibitem{Mourad} D.~Mourad, S.~Barthel, and G.~Czycholl, Phys.\ Rev.\ B {\bf 81}, 165316 (2010).

\bibitem{Harrison} F.~Long, W.E.~Hagston, P.~Harrison, T.~Stirner, J.\ Appl.\ Phys.\ {\bf 82}, 3414 (1997).

\bibitem{Zunger-far} A.~Zunger, phys.\ stat.\ sol.\ (a) {\bf 190}, 467 (2002).

\bibitem{Ivchenko_GX} Y.~Fu, M.~Willander, E.L.~Ivchenko, and A.A.~Kiselev, Phys.\ Rev.\ B {\bf 47}, 13498 (1993).

\bibitem{Krebs} O.~Krebs and P.~Voisin, Phys.\ Rev.\ Lett.\ {\bf 77}, 1829 (1996).

\bibitem{Ivchenko} E.L.~Ivchenko, A.Yu.~Kaminski, and U.~R\"ossler, Phys.\ Rev.\ B {\bf 54}, 5852 (1996).

\bibitem{Takhtamirov} E.E.~Takhtamirov and V.A.~Volkov, JETP {\bf  89}, 1000 (1999).

\bibitem{gamma-x} E.E.~Takhtamirov and V.A.~Volkov, JETP {\bf  90}, 1063 (2000).

\bibitem{stripes} E.E.~Takhtamirov and V.A.~Volkov, JETP Lett.\ {\bf 71} 422 (2000).

\bibitem{Luttinger} J.M.~ Luttinger and W.~Kohn, Phys.\ Rev.\ {\bf 97}, 869 (1955).

\bibitem{Luttinger2} J.M.~ Luttinger, Phys.\ Rev.\ {\bf 102}, 1030 (1956).

\bibitem{Leibler}  L.~Leibler, Phys.\ Rev.\ B {\bf 12}, 4443 (1975).

\bibitem{Zunger_large_good} A.~Zunger, phys.\ stat.\ sol.\ (b) {\bf 224}, 727 (2001).

\bibitem{Marquardt} O.~Marquardt, D.~Mourad, S.~Schulz, T.~Hickel, G.~Czycholl, and J.~Neugebauer, Phys.\ Rev.\ B {\bf 78}, 235302 (2008).

\bibitem{Grundmann} M.~Grundmann, O.~Stier, and D.~Bimberg, Phys.\ Rev.\ B {\bf 52}, 11969 (1995).

\bibitem{Stier} O.~Stier, M.~Grundmann, and D.~Bimberg, Phys.\ Rev.\ B {\bf 59}, 5688 (1999).

\bibitem{Andreev} A.D.~Andreev and E.P.~O'Reilly, Phys.\ Rev.\ B {\bf 62}, 15851 (2000).

\bibitem{Messiah} A.~Messiah, {\em Quantum Mechanics} (North-Holland, Amsterdam, 1961), Vol. I.

\bibitem{Landavshitz-QM}  L.D.~Landau and E.M.~Lifshitz, {\em Quantum Mechanics: Non-relativistic Theory}, 3nd ed. (Pergamon Press, Oxford, 1991).

\bibitem{SV} P.C.~Sercel and K.J.~Vahala, Phys.\ Rev.\ B {\bf 42}, 3690 (1990).

\bibitem{Baldereschi} A.~Baldereschi and N.O.~Lipari, Phys.\ Rev.\ B {\bf 8}, 2697 (1973).

\bibitem{Kane} E.O.~ Kane, J.\ Phys.\ Chem.\ Solids, {\bf 1}, 249 (1957).

\bibitem{Bir-Pikus} G.L.~Bir and G.E.~Pikus, {\em Symmetry and Strain-Induced Effects in Semiconductors}, (Wiley, New York, 1974).

\bibitem{Chuang} S.L.~Chuang and C.S.~Chang, Phys.\ Rev.\ B {\bf 54}, 2491 (1996).

\bibitem{Sirenko} Yu.M.~Sirenko, J.-B.~Jeon, K.W.~Kim, M.A.~Littlejohn, and M.A.~Stroscio, Phys.\ Rev.\ B {\bf 53}, 1997 (1996).

\bibitem{Ren} G.B.~Ren, Y.M.~Liu, and P.~Blood, Appl.\ Phys.\ Lett. {\bf 74}, 1117 (1999).

\bibitem{Landavshitz}  L.D.~Landau and E.M.~Lifshitz, {\em Theory of Elasticity}, 2nd ed. (Perga\-mon Press, Oxford, 1970).

\bibitem{Takhtamirov_sst} E.E.~Takhtamirov and V.A.~Volkov, Semicond.\ Sci.\ Technol. {\bf 12}, 77 (1997).

\bibitem{Korn} G.A.~Korn and T.M.~Korn, {\em Mathematical Handbook for Scientists and Engineers: Definitions, Theorems, and Formulas for Reference and Review}, (McGraw-Hill, New York, 1961).

\bibitem{Barettin} D.~Barettin, B.~Lassen, and M.~Willatzen, J.\ Phys.: Conf.\ Ser. {\bf 107}, 012001 (2008).

\bibitem{Frobenius} R.D.~Richtmyer, {\em Principles of Advanced Mathematical Physics} (Springer, Berlin, 1978), Vol.~1.

\bibitem{Bahder} T.B.~Bahder, Phys.\ Rev.\ B {\bf 41}, 11992 (1990).

\bibitem{Bahder_E} T.B.~Bahder, Phys.\ Rev.\ B {\bf 46}, 9913 (1992).

\bibitem{Vurgaftman_old} I.~Vurgaftman, J.R.~Meyer, and L.R.~Ram-Mohan, J.\ Appl.\ Phys. {\bf 89}, 5815 (2001).

\bibitem{Lowdin} P.-O.~L\"owdin, J.\ Chem.\ Phys. {\bf 19}, 1396 (1951).

\bibitem{Rinke} P.~Rinke, M.~Winkelnkemper, A.~Qteish, D.~Bimberg, J.~Neugebauer, and M.~Scheffler, Phys.\ Rev.\ B {\bf 77}, 075202 (2008).

\bibitem{Fu} J.Y.~Fu and M.W.~Wu, J.\ Appl.\ Phys. {\bf 104}, 093712 (2008).

\bibitem{Arley} M.~Arley, J.L.~Rouvi\`{e}re, F.~Widmann, B.~Daudin, G.~Feuillet, andH.~Mariette, Appl.\ Phys.\ Lett. {\bf 74}, 3287 (1999).

\bibitem{Bruls} D.M. Bruls, J.W.A.M.~Vugs, P.M.~Koenraad, H.W.M.~Salemink, J.H.~Wolter, M.~Hopkinson, M.S.~Skolnick, Fei~Long, and S.P.A.~Gill, Appl.\ Phys.\ Lett. {\bf 81}, 1708 (2002).

\bibitem{incorrectBC} A.V.~Maslov and C.Z.~Ning, Phys.\ Rev.\ B {\bf 72}, 125319 (2005).

\bibitem{globe1} M.~\c{S}ahin, S.~Nizamoglu, A.E.~Kavruk, and H.V.~Demir, J.\ Appl.\ Phys. {\bf 106}, 043704 (2009).

\bibitem{globe2} S.~Wu and L.~Wan, J.\ Appl.\ Phys. {\bf 111}, 063711 (2012).

\end{thebibliography}
\end{document}